\documentclass[a4paper,11pt]{article}

\usepackage{preamble}

\title{Optimal experimental design with k-space data: application to inverse hemodynamics}
\author{Miriam L\"ocke \and Ahmed Attia \and Dariusz Uci\'nski \and Crist\'obal Bertoglio}

\date{\today}

\begin{document}

\maketitle

\begin{abstract}

    Subject-specific cardiovascular models rely on parameter estimation using
    measurements such as 4D Flow MRI data. However, acquiring high-resolution,
    high-fidelity functional flow data is costly and taxing for the patient. As a
    result, there is growing interest in using highly undersampled MRI data to reduce
    acquisition time and thus the cost, while maximizing the information gain from the data.
    Examples of such recent work include inverse problems to estimate boundary conditions
    of aortic blood flow from highly undersampled k-space data.
    The undersampled data is selected based on a predefined sampling mask which
    can significantly influences the performance and the quality of the solution of
    the inverse problem.
    While there are many established sampling patterns to collect undersampled data,
    it remains unclear how to select the best sampling pattern for a given set of
    inference parameters.
    In this paper we propose an Optimal Experimental Design (OED) framework
    for MRI measurements in k-space, aiming to find optimal masks for estimating
    specific parameters directly from k-space. As OED is typically applied to sensor
    placement problems in spatial locations, this is, to our knowledge, the first time
    the technique is used in this context.
    We demonstrate that the masks optimized by employing OED consistently outperform conventional
    sampling patterns in terms of parameter estimation accuracy and variance, facilitating a
    speed-up of 10x of the acquisition time while maintaining accuracy.

\end{abstract}

\section{Introduction}
The personalization of hemodynamic models is a key aspect in achieving
patient-specific predictions in medical treatments. This requires high-quality clinical
data for the formulation and the numerical solution of inverse problems.
Using blood velocity measurements from Phase-Contrast Magnetic Resonance Imaging
(PC-MRI) is an appealing choice because it is both 
non-invasive and non-ionizing.
The acquisition time of 4D Flow MRI for a high-resolution image, however,
can be very high which can be taxing for both patients and medical
providers \cite{markl_4d_2012, van_schuppen_prerequisites_2025}.

As a result, it has become common practice to reduce the acquisition time by acquiring
only part of the frequency space (also called k-space) that MR images are measured in.
These undersampled measurements are then reconstructed by using Compressed Sensing (CS)
techniques \cite{sodhi_highly_2023,vargaszemes_highly_2023}.
Because MRI enables acquiring single lines of the
k-space, it provides a wide variety of possible undersampling patterns.
Different CS algorithms take advantage of the qualities of these patterns to construct accurate
images  \cite{neuhaus_accelerated_2019}.
For example, pseudo-random sampling patterns contain artifacts of an incoherent nature,
which can be removed by assuming sparsity of the true image in other domains.
Other CS algorithms leverage the mathematical
properties of other common designs such as spirals or radial patterns
\cite{negahdar_4_2016,kecskemeti_high_2012}, while others rely on broader properties
of the signal such as relative consistency in time or space  \cite{giese_highly_2013}.

Nonetheless, the reconstructed velocity images contain artifacts which negatively
impact the results of the solution of the inverse problem. Therefore, we have
recently proposed to solve the inverse problem directly from the undersampled frequency space
measurements \cite{locke_parameter_2025}.
This approach was exemplified via a Kalman filter adapted specifically for highly undersampled
k-space data to estimate boundary condition parameters of the underlying blood flow
model given by the incompressible Navier-Stokes equations.

This approach does not consider how the choice of sampling pattern depends on the
measurements and how it may affect the outcome of the inverse problem. Instead, there
is a plethora of possible design choices with barely even heuristics to make a
selection. Additionally, it seems reasonable that different boundary conditions or
flow parameters may be better estimated by different masks,
which has also not been explored.

Optimal Experimental Design
(OED)\cite{atkinson_optimum_2007,ucinski2004optimal,fedorov_model-oriented_2025} aims
to find optimal ways to perform an experiment, such as optimal sensor locations, based
on statistical criteria. Optimal designs can be used to reduce the costs of an
experiment, since a lesser number of measurements are needed to achieve the same
precision. Therefore, OED has been successfully used in applications in a large number
of different fields.

In this work we develop an OED approach to identify optimal
sampling patterns specific to the inference parameters, that is the parameters estimated
by the inverse problem.
We demonstrate that there are indeed specific frequencies that are
more informative for each parameter. Thus the OED-based optimal masks yield lower errors
and variances in the estimated parameters with a lower acquisition time due to the
reduced number of sampled frequencies.

The rest of the paper is organized as follows. \Cref{sec:theory} explains the
definitions of the measurement model, the OED problem, and the inverse problem.
\Cref{sec:analytic} describes the setup and results of a simplified analytical test
case, while \Cref{sec:aorta} provides the same for a more complex test case modelling
the hemodynamics in a section of the aorta. Finally, we conclude with a discussion of
our results in \Cref{sec:discussion} and a conclusion in \Cref{sec:conclusion}.

\section{Theory}\label{sec:theory}

\subsection{The forward problem}\label{subsec:model_dynamics}

Given the parameters $\bs{\theta}\in \mathbb{R}^p$, for example the physical constants
of PDE, and a dynamic state $\bs{X}^{k-1} \in \mathbb{R}^r$ at time $t^{k-1}$ with a
known initial condition $\bs{X}^{0}$, we define the forward problem as:
\begin{equation}
    \bs{X}^{k} = \mathcal{A}^k(\bs{X}^{k-1},\bs{\theta}), \quad k>0 \,,
\end{equation}
with $\mathcal{A}^k: \mathbb{R}^r \times \mathbb{R}^p \to \mathbb{R}^r$ mapping a
vector of parameters and the state from the time step $t^{k-1}$ to the state at
the next time step $t^{k}$.
Later, we will also use $\mathcal{A}^k(\bs{\theta})$ to
denote the solution of the forward problem at time $t^k$ based on the fixed initial
condition rather than a specific previous state, that is
\begin{equation}
    \bs{X}^{k} = \mathcal{A}^k \circ \cdots \cdot \circ \mathcal{A}^{1}
    (\bs{X}^{0},\bs{\theta}), \quad k>0 \,,
\end{equation}
where $\circ$ denotes the composition of operators.

We define the spatial undersampling operator as  $\mathcal{H}: \mathbb{R}^r \to
\mathbb{R}^{N_1 \times \dots \times N_D}$ for a number of dimensions $D$. For example,
for a two-dimensional test case $D = 2$, but for a full 3D acquisition of the blood
flow with a velocity vector consisting of components for each direction, $D = 4$ (one
    for each spatial dimension and
one for the velocity components). We will call $N = N_1\cdots N_D$ the total number of samples.

Next, we define the k-space observation operator $\mathcal{H}_{\mathcal{F}}:
\mathbb{R}^{r} \to \mathbb{R}^{N_1\times \dots \times N_D}$ as:
\begin{equation}
    \mathcal{H_{\mathcal{F}}}(\bs{X}) = \mathcal{F}\left[\bs{M}\odot
    e^{i\left(\frac{\pi}{venc}\mathcal{H}(\bs{X}) + \bs{\phi}_0\right)}\right] \,,
\end{equation}
where $\odot$ denotes the Hadamard product,
$\bs{X}$ is the dynamic state,
$venc$ is the velocity encoding parameter,
and $\mathcal{F}: \mathbb{C}^{N_1\times
\dots\times N_D} \to \mathbb{C}^{N_1\times \dots\times N_D}$ is the $D$-dimensional
discrete Fourier transform defined as
\begin{equation}
    \mathcal{F}[\bs{V}]_{y_1 \cdots y_D} = \sum_{n_1=0}^{N_1-1}\cdots\sum_{n_d =
    0}^{N_D - 1} \bs{V}_{n_1 \dots n_d} \exp\left(-i2\pi \left(\frac{y_1n_1}{N_1} +
    \cdots + \frac{y_Dn_d}{N_D}\right)\right) \,.
\end{equation}
The tensor $\bs{\phi}_0\in [-\pi, \pi)^{N_1\times \dots\times N_D}$ denotes the
background phase and $\bs{M} \in \mathbb{R}^{N_1\times \dots\times N_D}$ the magnitude
of the magnetization.
We assume that $\bs{\phi}_0$, $\bs{M}$, and $venc$ are all given.

For a \textit{true} parameter $\bs{\theta}_{true}$, 
the k-space
measurements at time $t^k$, i.e., $\bs{Y}^k \in \mathbb{C}^{N_1\times \cdots\times
N_D}$ for $k = 1, \dots, N_T$, are defined as:
\begin{equation}\label{eq:freq_meas}
    \bs{Y}^k =
    \left(\mathcal{H_{\mathcal{F}}}\left(\mathcal{A}^k(\bs{\theta}_{true})\right) +
    \bs{\epsilon}\right) \odot \bs{S} \,,
\end{equation}
where $\bs{\epsilon}\in \mathbb{C}^{N_1\times \dots\times N_D}$ is a complex Gaussian
noise with zero mean and standard deviation $\sigma_y$.
The measurement noise $\bs{\epsilon}$ is assumed to be proper,
meaning there are no correlations between the real and the imaginary parts of the signal.
The binary tensor $\bs{S}\in \mathbb{R}^{N_1\times \dots\times N_D}$ models the sampling mask
with a multi-index $i = (i_1, \dots, i_D)$:
\begin{equation}
    \bs{S}_{i} =
    \begin{cases}
        1 &\text{if the frequency $i$ is sampled}\,, \\
        0 & \text{otherwise} \,.
    \end{cases}
\end{equation}

The measurement provided in a \textit{zero-filled} form, with the
unavailable measurements replaced by $0$ such that the dimension of the undersampled
measurement matches the spatial
dimension despite the undersampling.
A key quantity for determining the scan time in MRI, when comparing the
accuracy given by different masks, is the acceleration factor $R$ defined as
\begin{equation}
    R = \frac{N}{\sum_{i} \bs{S}_{i}} \,,
\end{equation}
where $N$ is the total number of samples, and $\sum_{i} \bs{S}_{i}$ is
the total number of sampled frequencies.




\subsection{The inverse problem}\label{subsec:inverse_problem}



We formulate the parameter estimation problem as it is done in \cite{locke_parameter_2025}
through a data fidelity term in terms of the frequency space, therefore leading to the
minimization problem:
\begin{equation}
    \bs{\hat{\theta}}  \in  \argmin_{\bs{\theta} \in \mathbb{R}^p}
    \frac{1}{2\sigma_y^2} \sum_{k=1}^{N_T}\|\Re(\bs{Y}^k -
    \mathcal{H}_{\mathcal{F}}(\mathcal{A}^k(\bs{\theta}))\odot \bs{S})\|^2 +
    \|\Im(\bs{Y}^k - \mathcal{H}_{\mathcal{F}}(\mathcal{A}^k(\bs{\theta}))\odot \bs{S})\|^2
    \label{eq:freqJ}
\end{equation}
with the norm $\|\bs{X}\|^2 = \sum_{i_1, \dots, i_D} |\bs{X}_{i_1, \dots, i_D}|^2$.
Here $\Re(\cdot), \, \Im(\cdot)$, refer to the real- and the imaginary components of a
complex-valued signal.

In order to optimize the cost function, we are using the Reduced Order Unscented
Kalman Filter (ROUKF) to estimate the parameters $\bs{\theta}$ while assuming perfect
knowledge of the initial condition. ROUKF requires an initial guess for the
parameters. However, we perform some iterations on this by re-starting the ROUKF
algorithm using the result of the previous run as the new initial guess, while keeping
the standard deviation of the prior distribution unchanged, as was proposed in
\cite{pase_towards_2025}.
The details of the ROUKF algorithm is provided in \Cref{sec:app_ROUKF}.

This algorithm requires knowledge of the standard deviation $\sigma_y$ of the error.
As this is generally unknown, we instead estimate it by assuming that the signal in
the initial condition $\bs{X}^0$ is zero. Then the transformed signal in frequency
space at that time step should be $\mathcal{H_{\mathcal{F}}}(\bs{X}) =
\mathcal{F}\left[\bs{M}\odot
e^{i\bs{\phi}_0}\right]\odot \bs{S}$. The noise can now be isolated by subtracting
this quantity from the actual measurement, allowing the estimation of the standard deviation.

\subsection{The Optimal Experimental Design problem}

\subsubsection{Formulation}

The goal of the OED problem is to find the mask (the experimental design)
$\bs{S}$ out of a suitable design region
by solving the following optimization problem:
\begin{equation}\label{eqn:oed_opt}
    \bs{\hat{S}} = \argmin_{\bs{S}} \Phi(\bs{S}),
    \quad \text{s.t. }
    \sum_{i}\bs{S}_{i} = N_S \leq N \,,
\end{equation}
where $\Phi(\bs{S})$
is an OED optimality criterion that does not only depend on a given
mask $\bs{S}$, but also on the initial guess of the parameters $\bs{\theta}_0$ used
at the first ROUKF run, that is $\Phi(\bs{S}) \equiv \Phi(\bs{S},
\mathcal{H}_{\mathcal{F}}(\mathcal{A}(\bs{\theta}_0)))$.
%

Note that the constraint $\sum_{i}\bs{S}_{i} = N_S \leq N$ results in
a combinatorial problem of picking $N_S$ out of
$N$ voxels to construct a discrete design.

\subsubsection{Optimality criterion}

The most commonly used optimality criteria are based on the Fisher information matrix
(FIM) associated with the estimated parameters \cite{ucinski2004optimal,attia2018goal,
alexanderian_optimal_2021}.
Here, we construct the FIM based on the sensitivity coefficients, which are
derivatives of the observed output with respect to the unknown
parameters:
\begin{equation}
    \frac{\partial \widetilde{\bs{Y}}^k(\bs{\theta}_0)}{\partial \theta_1} , \dots,
    \frac{\partial \widetilde{\bs{Y}}^k(\bs{\theta}_0)}{\partial \theta_p}, \quad
    k=1,\dots,N_T \,,
\end{equation}
with
\begin{equation}
    \widetilde{\bs{Y}}^k(\bs{\theta}_0) =
    \mathcal{H}_{\mathcal{F}}(\mathcal{A}^{k}(\bs{\theta}_0))\odot \bs{S} \,,
\end{equation}
and therefore:
\begin{equation}
    \frac{\partial \widetilde{\bs{Y}}^k}{\partial \theta_j} = \mathcal{F}
    \left[i\frac{\pi}{venc} \bs{M} \odot
        e^{i(\frac{\pi}{venc}\mathcal{H}(\mathcal{A}^k(\bs{\theta})) + \bs{\phi}_0)} \odot
        \mathcal{H}\left(\frac{\partial}{\partial \theta_j} \mathcal{A}^k(\bs{\theta})\right)
    \right]\odot\bs{S}.
    \label{eq:sensitivities}
\end{equation}
The formula \eqref{eq:sensitivities} still requires the computation of the
sensitivities of the forward problem with respect to the parameters, which in this
work will be calculated
numerically using the finite difference method\cite[Chapter~2.6]{ucinski2004optimal}.
To be precise, we approximate
$\frac{\partial}{\partial\theta_j} \mathcal{A}^k(\theta)$ as
\begin{equation}
    \frac{\partial \mathcal{A}^k(\bs{\theta}) }{\partial\theta_j} \approx
    \frac{
        \mathcal{A}^k(\widetilde{\theta}) - \mathcal{A}^k(\theta)
    }{h\, \theta_j}
\end{equation}
where $\widetilde{\theta}$ is the disturbed vector of parameters such that
$\widetilde{\theta}_i = \theta_i$ for $i\neq j$, $\widetilde{\theta}_j = (1 + h) \theta_j$.

Finally, the FIM is defined by
\begin{equation}
    \bs{F} = \sum_{k = 1}^{N_T} \sum_{i = 1}^N \frac{2}{\sigma_y^2}
    \Re\left((\bs{G}^k)^*(\bs{G}^k)^T\right) \in \mathbb{R}^{p \times p}
\end{equation}
where $*$ is the complex conjugate transpose and $\bs{G}^k$ is the matrix of
sensitivity coefficients defined as
\begin{equation}
    \bs{G}^k =
    \begin{bmatrix}
        \displaystyle
        \left(vec\left(\frac{\partial \tilde{\bs{Y}}^k(\bs{\theta})}{\partial
        \theta_1}\right)\right)^{T} \\
        \vdots                                                           \\
        \displaystyle
        \left(vec\left(\frac{\partial \tilde{\bs{Y}}^k(\bs{\theta})}{\partial
        \theta_p}\right)\right)^{T}
    \end{bmatrix} \in \mathbb{C}^{p\times N} \,,
\end{equation}
with $vec(\bs{A})$ being the vector formed by stacking columns of the matrix $\bs{A}$.

There are several choices available for the design criterion $\Phi$ and it is not always clear
a priori which criterion leads to the best results. The most common choices, which we
will consider later in our numerical experiments, are \cite{fedorov_model-oriented_2025}:
\begin{enumerate}
    \item \textbf{A-criterion}: $\Phi = \text{tr}(\bs{F}^{-1})$. This criterion
        minimizes the trace of the inverse of the FIM, and therefore minimizes the
        average variance of the parameters.
    \item \textbf{D-criterion}: $\Phi = \det(\bs{F}^{-1})$. This criterion minimizes
        the volume of the uncertainty ellipsoid of the parameters, thus also maximizing
        the Shannon information content of the estimated parameters.
\end{enumerate}

\subsubsection{Optimization algorithms}

We consider two different algorithms to solve the combinatorial optimization problem
with binary designs.
The first is a greedy method, detailed in \Cref{alg:greedy}. At each
iteration the greedy algorithm chooses the frequency (out of all currently unused
frequencies) that minimizes the optimality criterion when added to the mask.
The algorithm stops when the specified number of frequencies has been selected. As the
FIM may be singular, a very small constant is added to the diagonal values to ensure
that inverting the matrix is possible.
The greedy algorithm does not guarantee finding the global optimum, but its time complexity is
linear in the number of frequencies $N_S$ to select and the total number of voxels
$N$, and therefore it is convenient in practice.

\begin{algorithm}[htbp!]
    \caption{Greedy algorithm}
    \begin{algorithmic}[1]
        \State Start with an empty mask $\bs{S}=[\bs{S}_i=0]_{i=(i_1, \dots, i_D)}$
        \State $j \gets 0$
        \While{$j < N_S$}
        \State Find 
        $k\!=\!(n_1, \dots, n_D)$ that
        minimizes $\Phi(\bs{S})$
        when setting $\hat{\bs{S}}_k = 1$ where $\bs{S}_k = 0$
        \State Set $\bs{S}_k = 1$
        \State $j \gets j + 1$
        \EndWhile
    \end{algorithmic}
    \label{alg:greedy}
\end{algorithm}

The second algorithm we consider is the exchange method \cite{fedorov_model-oriented_2025},
detailed in \Cref{alg:exchange}.
The method starts with a random selection of the
specified number of frequencies $N_s$.
At each iteration, the algorithm finds the best
exchange between any one of the currently selected frequencies
and a neighboring frequency that is not currently part of the design.
The neighboring indices are defined as being contained within a sphere of a given radius
around the currently examined frequency.
For a sufficiently large radius, the algorithm will
therefore consider exchanges with all available frequencies.
This algorithm terminates
if there is no exchange possible that improves the optimality criterion beyond a
certain tolerance.
The total runtime of the algorithm is not known a priori and also
depends on the random seed used to generate the initial mask.

A critical factor influencing the runtime of the exchange algorithm is possibly
frequent inversion of the FIM after perturbation of the original FIM. That is why in
the analytical case below the Sherman-Morrison-Woodbury formulae have been used.
Indeed, it is easy to see that addition/deletion of a frequency to/from the current
set of selected frequencies amounts to a rank-2 perturbation of the FIM. This
approach, however, cannot be applied to the aortic hemodynamics setting that follows,
since the appropriate perturbations have ranks $\min(p, 2 N_T)$, i.e., they are by no
means small-rank. Therefore, relatively simple exchanges typically used for numerical
construction of directly constrained design measures
\cite[Chapter~4.3]{fedorov_model-oriented_2025}, also known as clusterization-free
designs\cite[Chapter~3.4]{ucinski2004optimal}(Uciński, 2005, Ch. 3.4) have been
adopted. The technique is applicable in the case of a dense grid of frequencies, which
is valid here.

\begin{algorithm}[htbp!]
    \caption{Restricted exchange algorithm}
    \begin{algorithmic}[1]
        \State Sample $N_S$ frequencies randomly to create $\bs{S}^0$
        \State $k\gets 1$
        \While{True}
        \State Define the neighborhood $\mathcal{N}\left(\bs{S}^{k-1}\right) = \{i
        \text{ s.t.} ||k-i||<radius \text{ for a } j\in \bs{S}^{k-1}, \bs{S}^{k-1}_j = 1 \}$
        \State Find $(i^k, j^k) \in \bs{S}^{k-1} \times\left(
        \mathcal{N}\left(\bs{S}^{k-1}\right)\setminus\bs{S}^{k-1}\right)$  that
        minimizes $\Psi(\bs{S}^{k-1}_{i\rightleftarrows j}$ where
            $\bs{S}^{k-1}_{i\rightleftarrows j}$ switches the values of $\bs{S}$ at $i$ and $j$.
            \If{$\Psi(\bs{S}^{k-1}_{i\rightleftarrows j}) < \Psi(\bs{S}^{k-1})$}
            \State Set $\bs{S}^k = \bs{S}^{k-1}_{i\rightleftarrows j}$ and continue
            \Else
            \State \textbf{break}
            \Comment{stop if there is insufficient improvement}
            \EndIf
            \EndWhile
        \end{algorithmic}
        \label{alg:exchange}
    \end{algorithm}

    \section{Analytical Test Case}\label{sec:analytic}
    In this section we consider a simple case in which the measurement sensitivities
    can be computed analytically. Additionally, we construct the signal to have a
    specific significant frequencies in order to be able to confirm the results of the
    OED algorithms. First, we therefore describe the forward problem and our
    measurements, then the computation of the sensitivities, followed by details of
    the implementation of the algorithms. We then show both the optimal masks and the
    conventional masks we are using for comparison and their results on the inverse
    problem. Lastly, we explore the effects of using sensitivities computed with
    approximated parameters rather than accurate ones.

    \subsection{Forward and measurement models}

\begin{figure}[htbp!]
        \centering
        \subfloat[Function $f(x,y)$]{
        \includegraphics[width=0.33\textwidth]{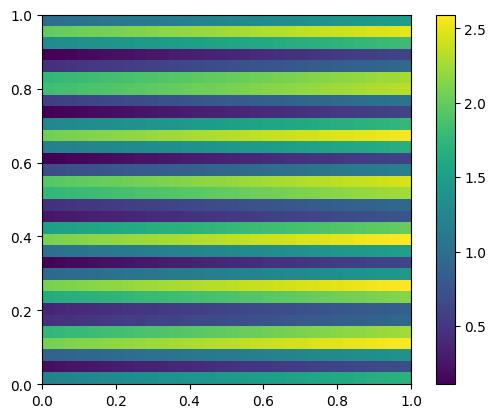}}
        \subfloat[Measurements (Real part)]{
        \includegraphics[trim=50 10 50 10, clip, width=0.33\textwidth]{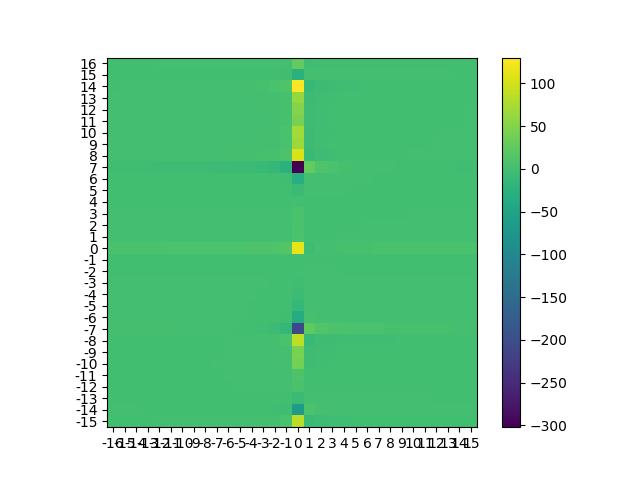}}
        \subfloat[Measurements (Imaginary part)]{
            \includegraphics[trim=50 10 50 10, clip,
            width=0.33\textwidth]{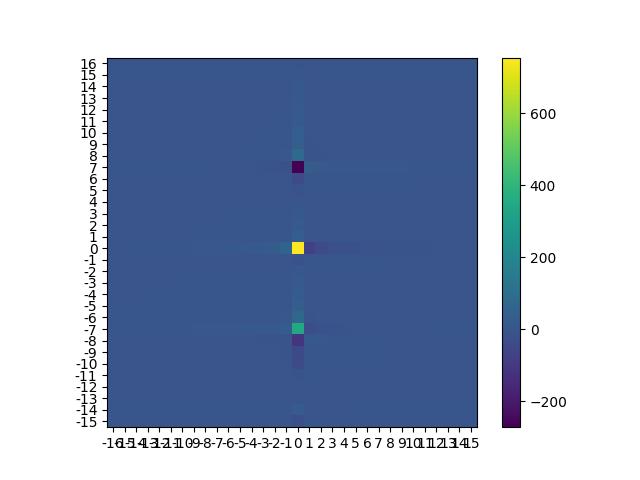}
        }
        \caption{Visualization of the analytical function \eqref{eqn:analytical_fun}  (a),
        and the MRI-like measurement signal in frequency space (b) and (c).}
        \label{fig:polynomial_images}
    \end{figure}

    To explore the effect of specific frequencies, we use a simple two-dimensional function
    \begin{equation}\label{eqn:analytical_fun}
        f(x,y) = c_0 + c_1 x + c_2 y + c_3 \sin(2\pi\omega x) \,,
    \end{equation}
    with $c_0 = 1.0, c_1 = 0.2, c_2 = 0.5, c_3 = 1.0$
    which combines a first-order polynomial with a sine function with a frequency $\omega
    = 7$. The coefficients $c_i$ are the parameters to be estimated, while $\omega$ is
    assumed to be known. We consider this function in the domain $[0, 1]^2$ which is
    discretized with $32$ elements in each direction, which in frequency space results in
    $16$ positive and $15$ negative frequencies in each direction.
    The function is transformed into MRI-like frequency space data as described
    by \eqref{eq:freq_meas}, by using a constant magnitude of $1$,
    a constant background phase $\phi_0 = 0.0075$ and velocity encoding parameter
    $venc = 3$. %
    \Cref{fig:polynomial_images} shows images of the original function as well
    as the frequency space data. In frequency space, despite the MRI-like transform
    preceding the Fourier transform, the sinusoidal frequency at $\omega = 7$ is clearly
    visible, as well as the horizontal and vertical ``stripes" deriving from the
    convolution with the linear parts of the function.

    \subsection{Sensitivities}

\begin{figure}[htbp!]
        \centering
        \subfloat[$c_0$]{
        \includegraphics[width=0.25\textwidth]{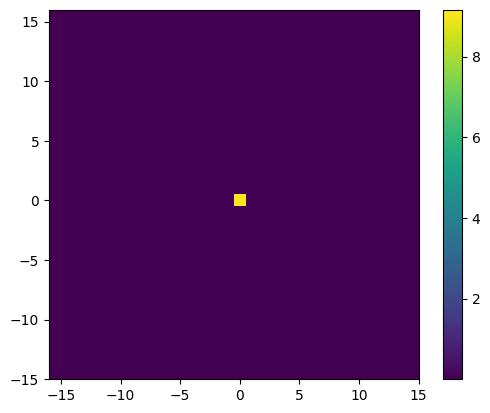}}
        \subfloat[$c_1$]{
        \includegraphics[width=0.25\textwidth]{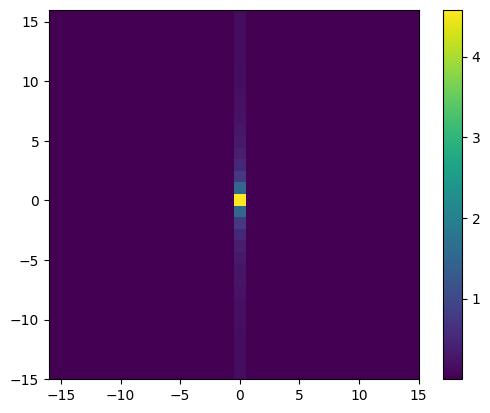}}
        \subfloat[$c_2$]{
        \includegraphics[width=0.25\textwidth]{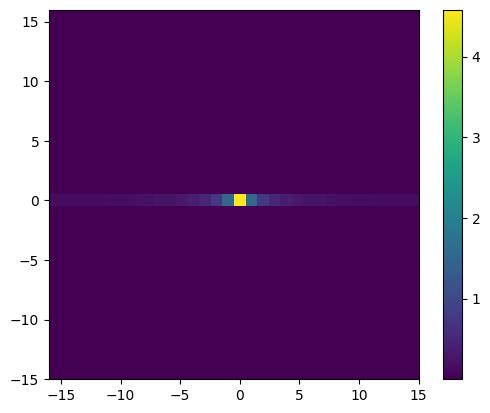}}
        \subfloat[$c_3$]{
        \includegraphics[width=0.25\textwidth]{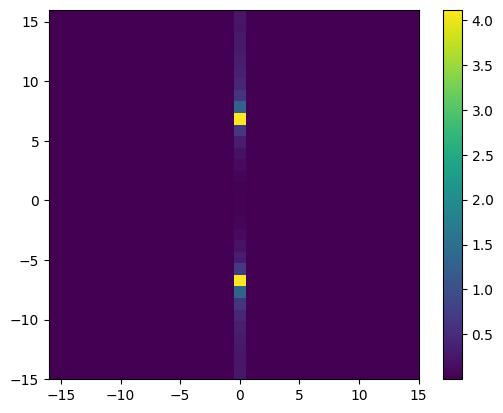}}
        \caption{Magnitude of the sensitivities of the measurement for each of the
            parameters in the analytical test case \eqref{eqn:analytical_fun}.
            The parameters here are
            the constant component $c_0$,
            the linear x-component $c_1$,
            linear y-component $c_2$, and
            the sinusoidal component $c_3$, respectively.
        }
        \label{fig:polynomial_sens}
    \end{figure}

    The sensitivities for each parameter are computed analytically according to 
    \eqref{eq:sensitivities}. Images of the magnitudes of the sensitivities for each
    parameter are shown in \Cref{fig:polynomial_sens}. It is apparent that different
    frequencies are more sensitive to each of the coefficients, depending on their basis
    function. Notably, as $c_0, c_1$ and $c_2$ all relate to polynomials, the highest
    values of their sensitivity is concentrated around the center frequency $(0, 0)$,
    while this frequency has a sensitivity close to zero for $c_3$, which relates to a
    sinusoidal function.

    \subsection{Implementation of the optimization algorithms}

    For the greedy algorithm, we use the implementation provided by the PyOED package
    \cite{chowdhary_pyoed_2024}.
    For the exchange algorithm, we choose to use a radius of 10 voxels and a
    tolerance of $10^{-8}$, as we have found that further increasing the radius or
    decreasing the tolerance does not change the result of the algorithm. We use our own
    implementation of this algorithm in Python.

    In terms of computational cost, we list the computation times in
    \Cref{tab:timing_analytical}, computed on a notebook with 16GB RAM and an Intel Core i5
    processor. The greedy algorithm requires a longer computation time than the exchange
    algorithm for lower budgets, but scales better, leading to a shorter time for a budget
    of 50 points.
    \begin{table}[htbp!]
        \centering
        \begin{tabular}{|c|c|c|c|c|}
            \hline
            Algorithm & \multicolumn{4}{c|}{budget ($N_s$)}       \\ \cline{2-5}
            & 5                           & 10     & 25     & 50       \\
            \hline
            Greedy    & 6.5s                        & 12.96s & 33.81s & 1min 27s \\
            Exchange  & 0.49s                       & 3.09s  & 18.23s & 1min 6s \\
            \hline
        \end{tabular}
        \caption{Compute time (in seconds) for the two algorithms for different budgets on the
        analytical testcase \eqref{eqn:analytical_fun}.}
        \label{tab:timing_analytical}
    \end{table}

    \subsection{Optimal masks}

\begin{figure}[htbp!]
        \centering
        \subfloat[A-opt; Greedy]{
        \includegraphics[width=0.25\textwidth]{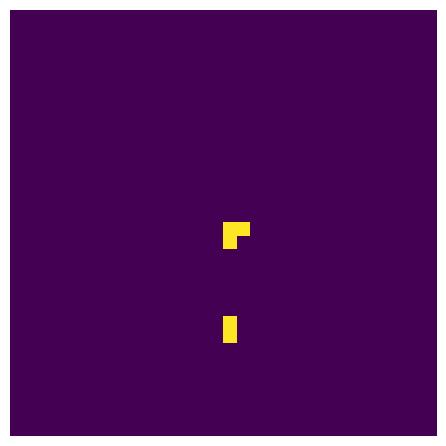}}
        \subfloat[A-opt; Exchange]{
            \includegraphics[width=0.25\textwidth]{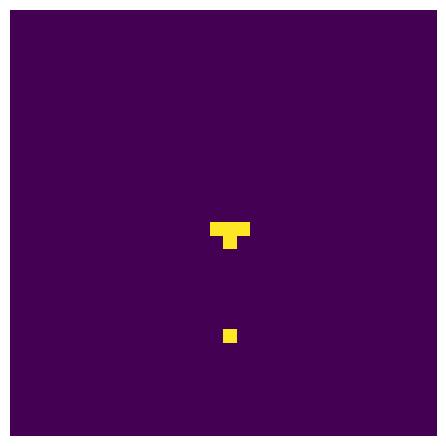}
        }
        \subfloat[D-opt; Greedy]{
            \includegraphics[width=0.25\textwidth]{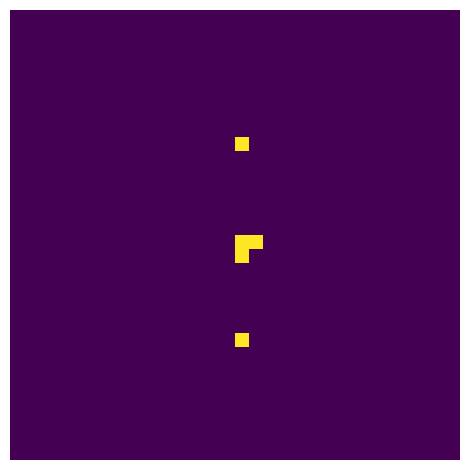}
        }
        \subfloat[D-opt; Exchange]{
            \includegraphics[width=0.25\textwidth]{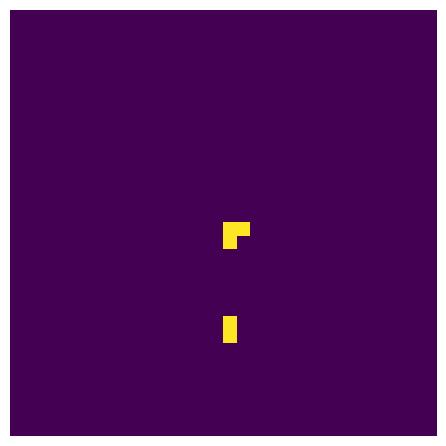}
        }
        \caption{Masks for a budget of $N_S = 5$ frequencies.}
        \label{fig:polynomial_budget_5}
    \end{figure}

\begin{figure}[htbp!]
        \centering
        \subfloat[A-opt; Greedy]{
        \includegraphics[width=0.25\textwidth]{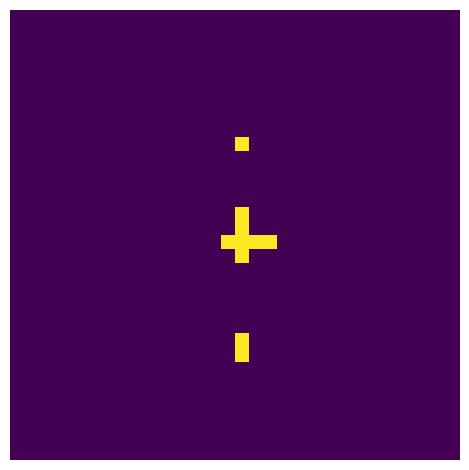}}
        \subfloat[A-opt, Exchange]{
            \includegraphics[width=0.25\textwidth]{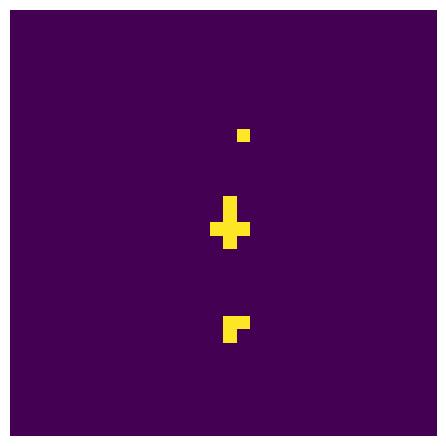}
        }
        \subfloat[D-opt, Greedy]{
            \includegraphics[width=0.25\textwidth]{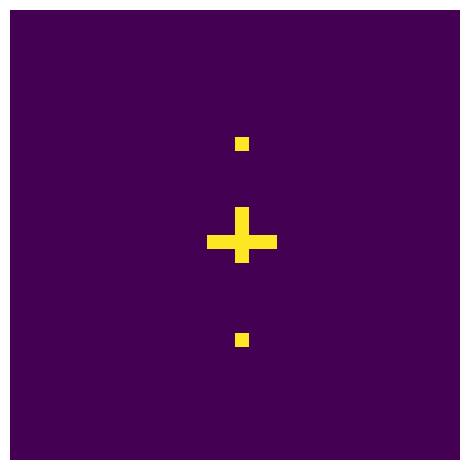}
        }
        \subfloat[D-opt, Exchange]{
            \includegraphics[width=0.25\textwidth]{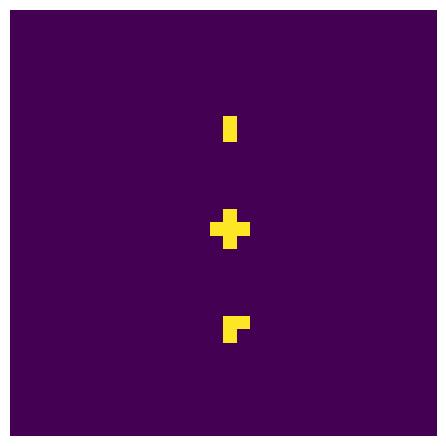}
        }
        \captionof{figure}{Masks for a budget of $N_S = 10$ frequencies.}
        \label{fig:polynomial_budget_10}
    \end{figure}

\begin{figure}[htbp!]
        \centering
        \subfloat[A-opt, Greedy]{
        \includegraphics[width=0.25\textwidth]{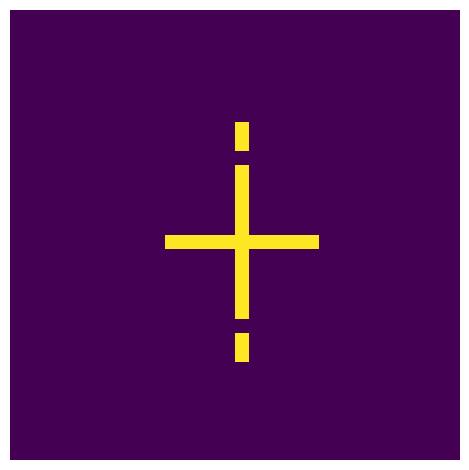}}
        \subfloat[A-opt, Exchange]{
            \includegraphics[width=0.25\textwidth]{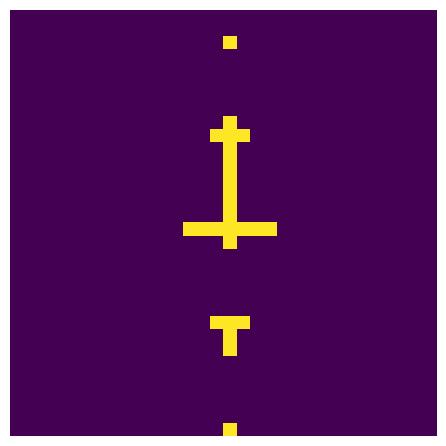}
        }
        \subfloat[D-opt, Greedy]{
            \includegraphics[width=0.25\textwidth]{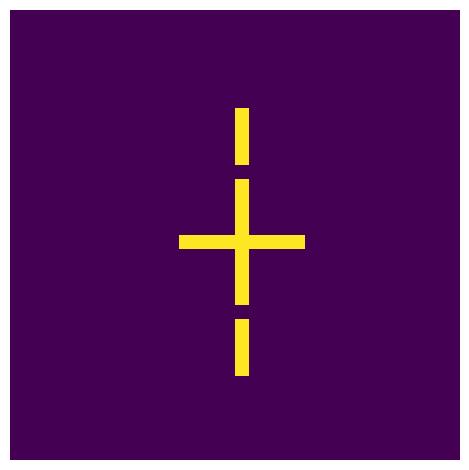}
        }
        \subfloat[D-opt, Exchange]{
            \includegraphics[width=0.25\textwidth]{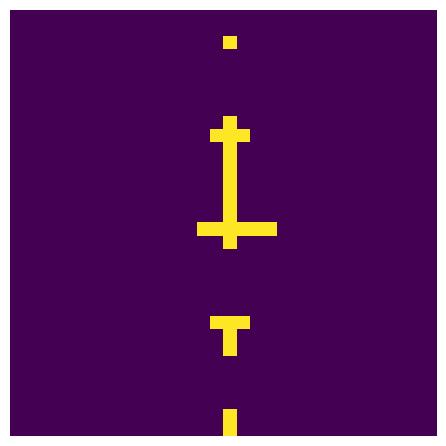}
        }
        \caption{Masks for a budget of $N_S = 25$ frequencies.}
        \label{fig:polynomial_budget_25}
    \end{figure}

\begin{figure}[htbp!]
        \centering
        \subfloat[A-opt, Greedy]{
        \includegraphics[width=0.25\textwidth]{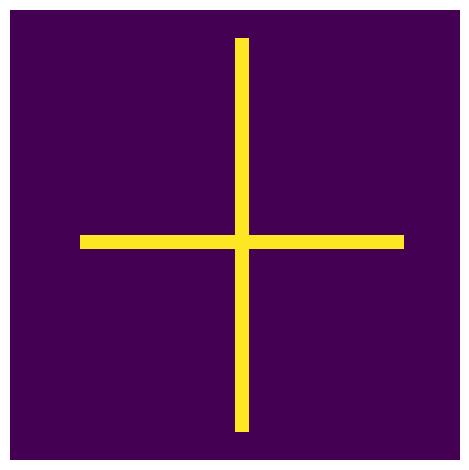}}
        \subfloat[A-opt, Exchange]{
            \includegraphics[width=0.25\textwidth]{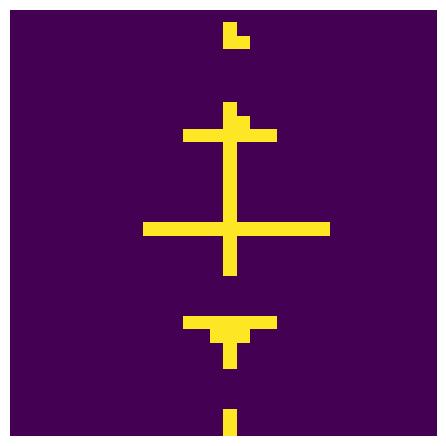}
        }
        \subfloat[D-opt, Greedy]{
            \includegraphics[width=0.25\textwidth]{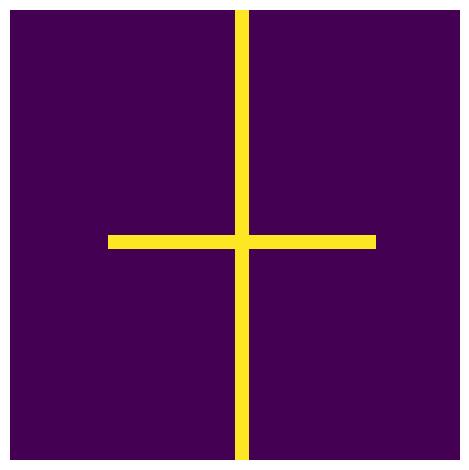}
        }
        \subfloat[D-opt, Exchange]{
            \includegraphics[width=0.25\textwidth]{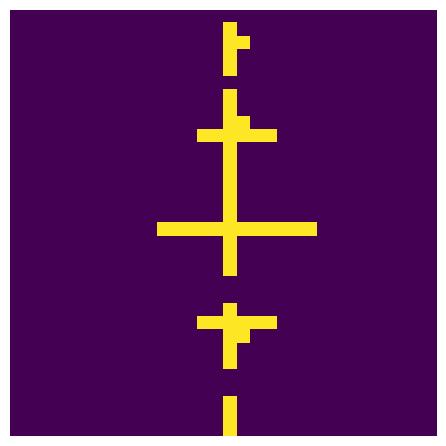}
        }
        \caption{Masks for a budget of $N_S = 50$ frequencies.}
        \label{fig:polynomial_budget_50}
    \end{figure}

    We solve the OED problem with both A-optimality and D-optimality criteria, and with
    masks for four different budgets $N_S = 5, 10, 25, 50$. These budgets correspond
    to acceleration factor values of $R \sim 205, 102, 41, 21$, respectively.
    The resulting optimal masks for the combinations of optimality criteria, budgets, and
    optimization algorithms, are shown in Figures
    \ref{fig:polynomial_budget_5}---\ref{fig:polynomial_budget_50}. 

    For lower budgets, as seen in Figures \ref{fig:polynomial_budget_5} and
    \ref{fig:polynomial_budget_10}, the masks are very similar between criteria and
    algorithms, with frequencies near the center and at the wave frequency mainly being selected.

    For higher budgets, as seen in Figures \ref{fig:polynomial_budget_25} and
    \ref{fig:polynomial_budget_50}, differences between the algorithms become apparent.
    While the greedy algorithm tends to select frequencies mainly along the vertical and
    horizontal axes, the exchange algorithm clusters points around the center and the wave
    frequency. For both algorithms, the D-criterion results in more points being selected
    along the vertical axis, whereas the A-criterion leads to the selection of more
    horizontal points, either on the central axis or as ``branches" off points on the
    central vertical axis.

    \subsection{Conventional masks}
\begin{figure}[htbp!]
        \centering
        \subfloat[Gaussian; $N_s=5$]{
        \includegraphics[width=0.25\textwidth]{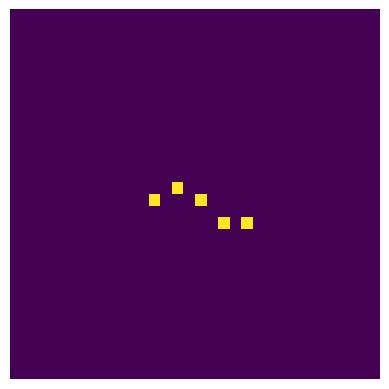}}
        \subfloat[Gaussian; $N_s=10$]{
        \includegraphics[width=0.25\textwidth]{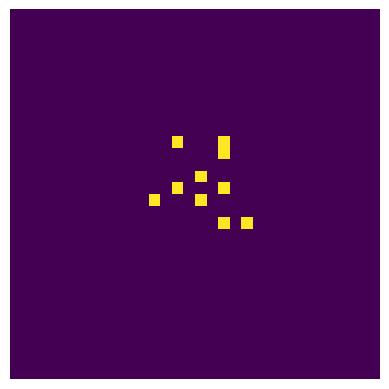}}
        \subfloat[Gaussian; $N_s=25$]{
        \includegraphics[width=0.25\textwidth]{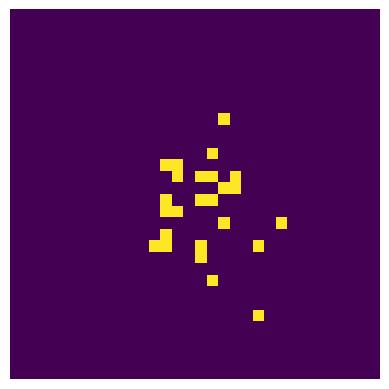}}
        \subfloat[Gaussian; $N_s=50$]{
        \includegraphics[width=0.25\textwidth]{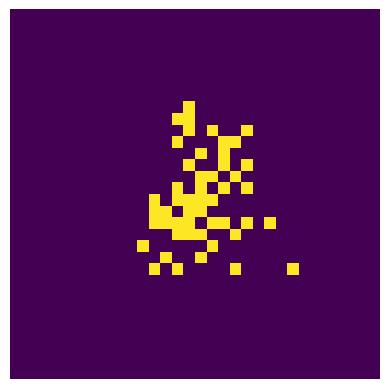}}\\
        \subfloat[Circle; $N_s=5$]{
        \includegraphics[width=0.25\textwidth]{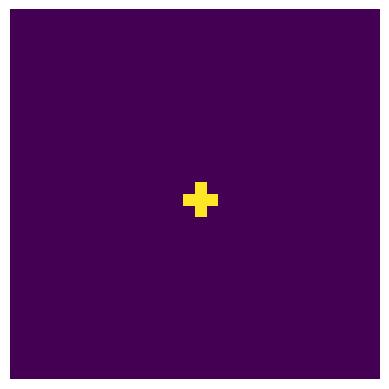}}
        \subfloat[Circle; $N_s=$ (9 points)]{
        \includegraphics[width=0.25\textwidth]{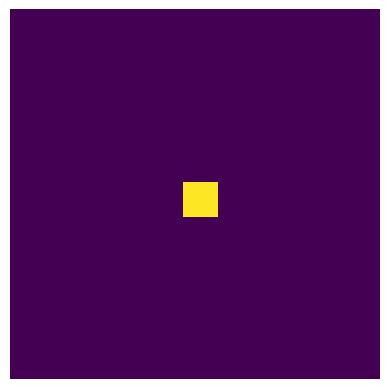}}
        \subfloat[Circle; $N_s=25$]{
        \includegraphics[width=0.25\textwidth]{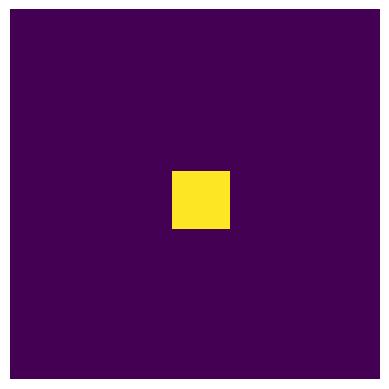}}
        \subfloat[Circle; $N_s=50$ (49 points)]{
        \includegraphics[width=0.25\textwidth]{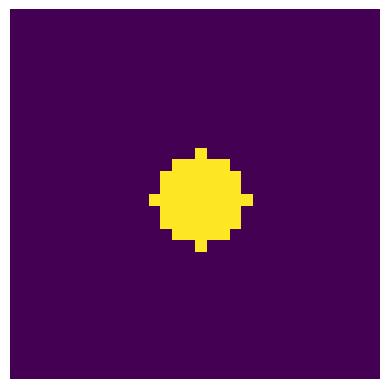}}
        \caption{Masks generated using a pseudo-random Gaussian sampling pattern
            and a circular sampling pattern, for a variety of acceleration factors in
            the analytical
            test case \eqref{eqn:analytical_fun}.
        }
        \label{fig:conv_masks}
    \end{figure}

    The OED-based masks are compared here with some of the conventional undersampling masks;
    see e.g., \cite{locke_parameter_2025}.
    Specifically, we show a pseudo-random Gaussian pattern as well as a
    ``circle" pattern that consists of a circle positioned in the middle of the frequency
    space.
    Due to the constraints of this design, it was not always possible to match the
    budgets exactly, but they are off by at most one point.
    Hence the circular masks also
    occasionally appear more in the shape of a square. The resulting masks are shown in
    \Cref{fig:conv_masks}.

    \subsection{Parameter estimation}

    We generate thirty realizations of Gaussian noise added to undersampled k-space
    measurements such that the signal-to-noise ratio (SNR) is $15$, amounting to
    $\sigma_y\approx 5.69$. The parameters are
    estimated by using ROUKF as discussed in \Cref{subsec:inverse_problem}.
    Because the polynomial
    is not time-dependent, only a single step is performed; however, we iterate the
    estimation ten times by providing the result of the previous iteration as the new
    initial parameter value for the next iteration of the filter.
    The true solution is provided to the filter as the initial guess. The standard
    deviation of the noise $\sigma_y$ is provided to the filter in this case as well,
    as there is no frame with close-to-zero signal to estimate it from.

\begin{figure}[htbp!]
        \centering
        \includegraphics[width=0.6\textwidth]{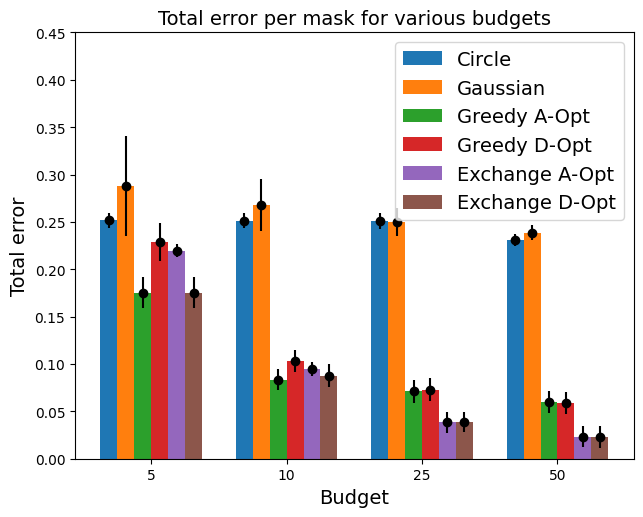}

        \caption{Total error
            $e = \frac{||\bs{\theta} - \bs{\theta}_{true}||}{||\bs{\theta}_{true}||}$
            of the estimated parameters $\bs{\theta}$ 
            by employing the
            A- and the D-optimal masks, as well as the circle and the Gaussian masks.
            Results are shown for
            various budgets $N_s$ for the analytical test case \eqref{eqn:analytical_fun}.
        }
        \label{fig:polynomial_errors}
    \end{figure}

    \Cref{fig:polynomial_errors} shows the quality of the inference results in
    terms of the relative error of the estimated parameter $\bs{\theta}$
    with respect to the true parameter values $\bs{\theta}_{true}$, that is
    $e = \frac{||\bs{\theta} - \bs{\theta}_{true}||}{||\bs{\theta}_{true}||}$.
    It can be seen that the optimal designs perform considerably better than the
    conventional designs, even outperforming conventional designs with a higher budget.
    The variance is also noticeably lower for the optimal masks. There appear to be only
    minor differences between the two criteria (well within the variance), though the
    A-optimal masks achieve lower error values than the D-optimal masks in some cases.
    While the differences between the two algorithms are minor for the higher budgets, for
    lower budgets it can be seen that the exchange algorithm results in decreased errors
    and variances for both optimality criteria.

\begin{figure}[htbp!]
        \centering
        \subfloat[$c_0$, constant component]{
            \includegraphics[width=0.5\textwidth]{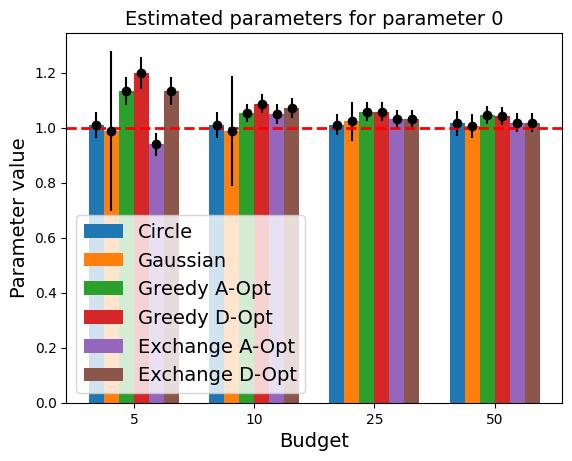}
        }
        \subfloat[$c_1$, x-linear component]{
            \includegraphics[width=0.5\textwidth]{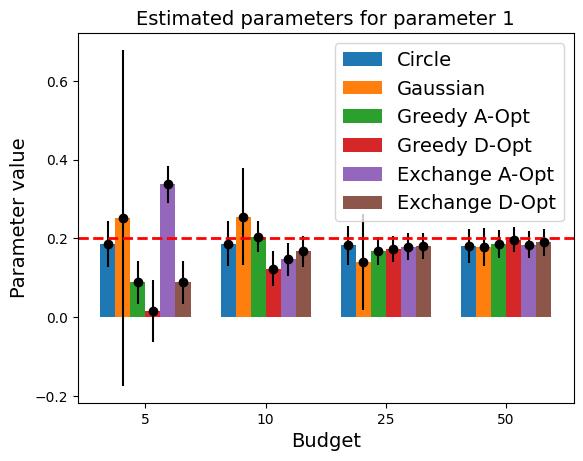}
        }\\
        \subfloat[$c_2$, y-linear component]{
            \includegraphics[width=0.5\textwidth]{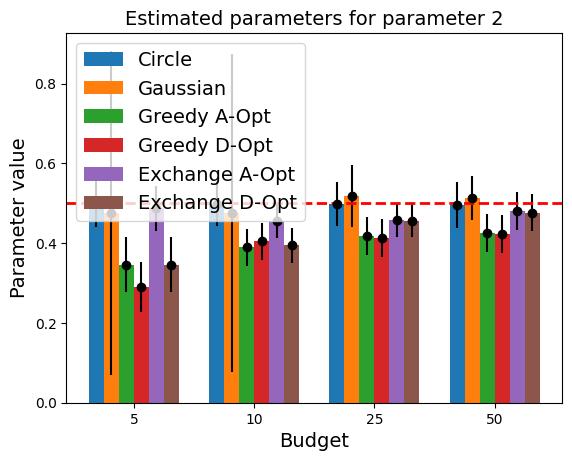}
        }
        \subfloat[$c_3$, sinusoidal component]{
            \includegraphics[width=0.5\textwidth]{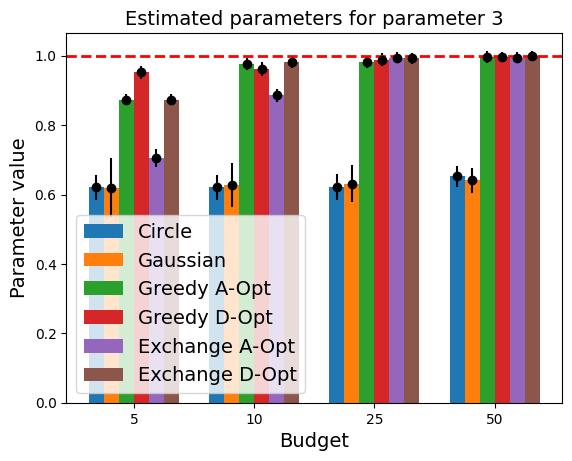}
        }
        \caption{Estimated values of the parameters for different masks averaged over
            the noise realizations. Red dashed lines
            indicate true values of the parameters in the analytical test case
            \eqref{eqn:analytical_fun}. For
        visibility, the variance indicated by the black bars has been multiplied by 3.}
        \label{fig:polynomial_param_bars}
    \end{figure}

    In \Cref{fig:polynomial_param_bars}, the estimated values of each coefficient of
    the function are displayed for the different masks and budgets. It can be seen that
    the variance is noticeably lower for the optimal masks compared to the
    Gaussian masks, and comparable or slightly lower compared to the circle masks. The
    largest variances can be observed for parameter 1, and the lowest for parameter 3.
    This appears to be consistent between the masks.

\begin{figure}[htbp!]
        \centering
        \subfloat[A-optimality criterion]{
            \includegraphics[width=0.5\textwidth]{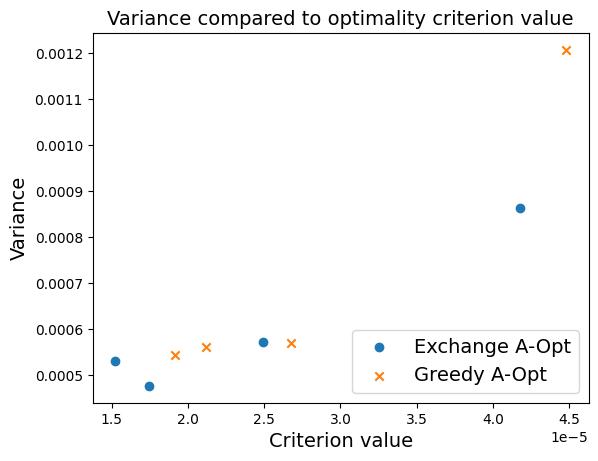}
        }
        \subfloat[D-optimality criterion]{
            \includegraphics[width=0.5\textwidth]{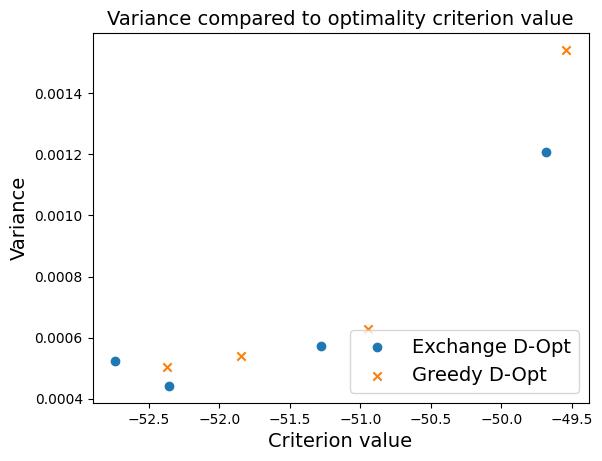}
        }

        \caption{Optimality criterion values compared to the sum of the variances of the
            estimated parameters in the analytical test case \eqref{eqn:analytical_fun}.
            A lower optimality criterion
            value is better.
        }
        \label{fig:polynomial_crit_vars}
    \end{figure}

    We plot the sum of the variances of the estimated parameters in
    \Cref{fig:polynomial_crit_vars}. As expected, this shows good agreement between a lower
    criterion value and a lower variance.
    The differences in the performance of the different masks are noticeable, with the
    conventional mask performing better for the constant and linear components for lower
    budgets, but highly underestimating the coefficient of the sine function even for the
    highest budget. This is likely because neither the circle nor the Gaussian includes
    the $k = (0, \pm 7)$ frequency, which has the highest sensitivity for that parameter.
    Conversely, the optimal masks estimate this parameter well even for lower budgets.
    Therefore there are less points included in these masks that are relevant for the
    constant and linear components, thus there is a higher error for these parameters for
    the low budgets.

\begin{figure}[htbp!]
        \centering
        \subfloat[A-optimality criterion]{
            \includegraphics[width=0.5\textwidth]{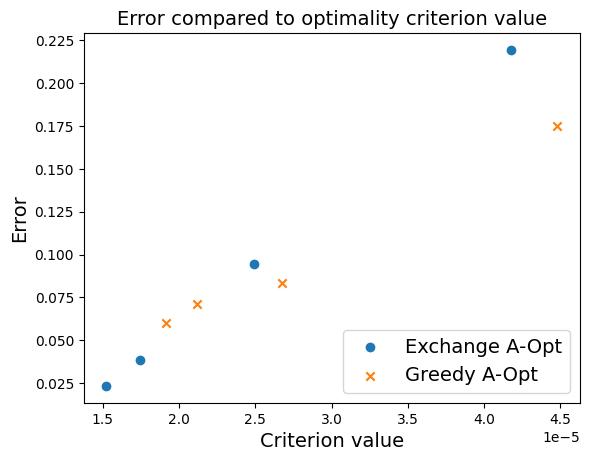}
        }
        \subfloat[D-optimality criterion]{
            \includegraphics[width=0.5\textwidth]{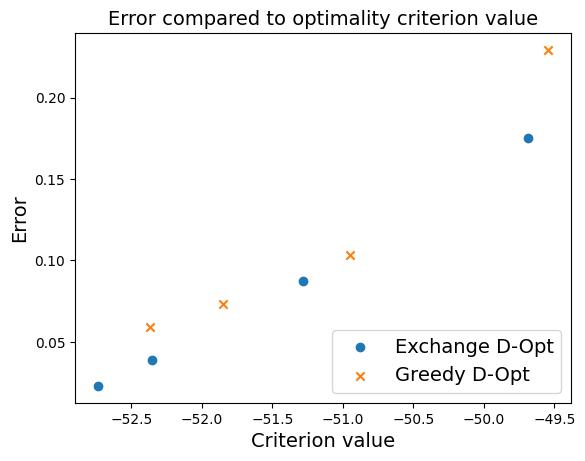}
        }

        \caption{Optimality criterion values compared to the errors in the analytical test case
        \eqref{eqn:analytical_fun}. A lower optimality criterion value is better.}
        \label{fig:polynomial_crit_values}
    \end{figure}

    It is worth noting that none of the optimality criteria optimize directly for the
    accuracy of the estimated parameters. In \Cref{fig:polynomial_crit_values}, we
    therefore plot the optimality criterion values against the error for each set of
    optimal masks and the conventional masks. It can be seen that there is a strong trend:
    a better criterion value (lower for both the A- and D-criteria) generally leads to a
    lower error, especially for the lower budgets. For higher budgets, there are
    occasional outliers where a higher criterion value nonetheless corresponds to a lower
    error. It can also be seen that the exchange algorithm achieves better criterion
    values than the greedy algorithm.

    \subsection{Using sensitivities for estimated parameter values}

    So far, we have calculated the sensitivities using the true parameters. However,
    generally these values will not be available. Therefore we now consider sensitivities
    computes with incorrect parameter values, those being $c_0 = c_1 = c_2 = c_3 = 0.1$.

\begin{figure}[htbp!]
        \centering
        \subfloat[exchange algorithm]{
        \includegraphics[width = 0.5\textwidth]{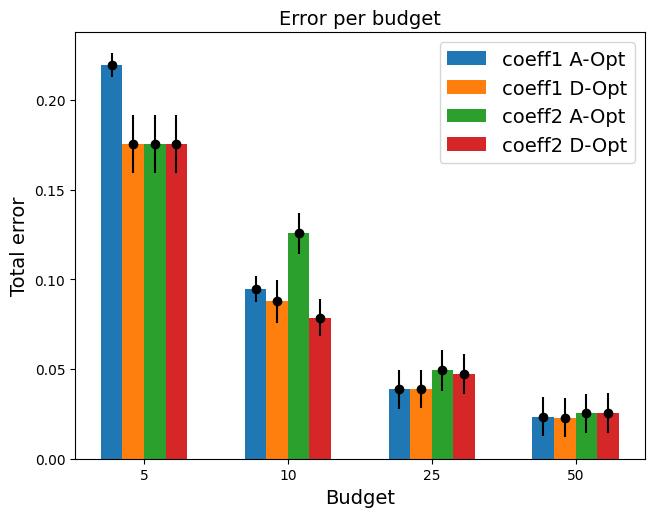}}
        \subfloat[greedy algorithm]{
        \includegraphics[width = 0.5\textwidth]{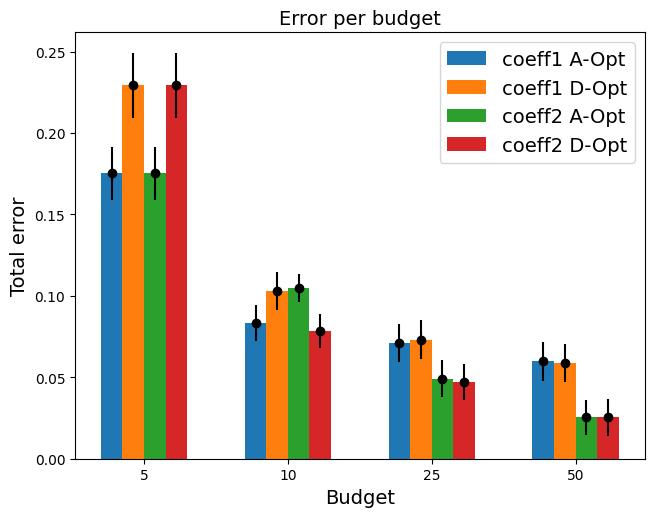}}
        \caption{Comparison of the error of the inverse problem between the sensitivities
            computed with two different sets of coefficients, \texttt{coeff1} = $[1.0, 0.2, 0.5,
            1.0]$, which are the true values, and \texttt{coeff2} = $[0.1, 0.1, 0.1, 0.1]$.
            \label{fig:polynomial_approx_sens}
        }
    \end{figure}

    The optimal masks computed with these sensitivities generally often coincide with at
    least one mask computed with the true sensitivities, and differ only slightly
    otherwise. The results of the inverse problem, for the true and approximated
    sensitivities, are shown in \Cref{fig:polynomial_approx_sens}. It can be seen
    that there are rarely significant differences due to the sensitivities.

    As can be seen from these results, using masks computed by OED algorithms,
    regardless of criterion, identify the significant frequencies and provide large
    improvements in terms of error and variance in the parameters estimated by the
    inverse problem. We will therefore now test our approach in a more realistic application.

    \section{Aortic Hemodynamics Test Case}\label{sec:aorta}

    \subsection{Forward model}

\begin{figure}[hbtp!]
        \centering
        \includegraphics[width=0.37\textwidth]{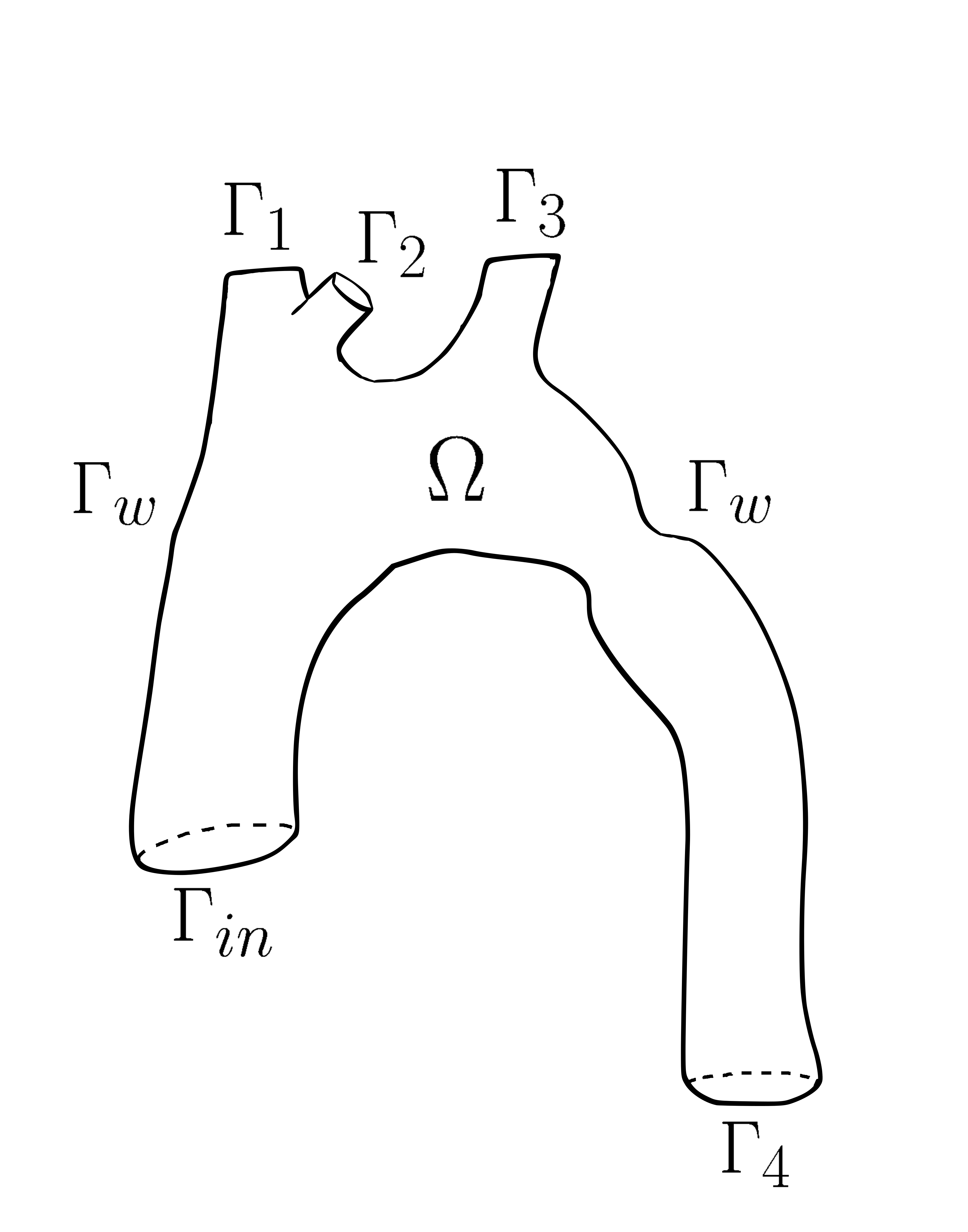}
        \caption{3D aortic model geometry}
        \label{fig:aorta}
    \end{figure}

    We use the same model and process as is used in \cite{locke_parameter_2025}.  We
    consider a geometry of the lumen of the ascending and descending aorta including the
    outlets of the brachycephalic artery, left common carotid artery, and left subclavian
    artery, as depicted in \Cref{fig:aorta}. This geometry serves as the domain for
    the forward model. The geometry was discretized with unstructured trapezoidal elements
    with a total of 20,916 points.
    The boundary of the geometry consists of six different boundaries: $\Gamma_{in}$ being
    the inlet boundary in the ascending aorta, $\Gamma_{w}$ the arterial wall, and the
    remaining boundaries $\Gamma_l$ for $l = 1,\dots,4$ representing the outlets.

    We model the blood flow in this domain with the incompressible Navier-Stokes equations
    for the velocity $\bs{u}(\bs{x}, t)\in \mathbb{R}^3$ and the pressure $p(\bs{x},t)\in
    \mathbb{R}$:
    \begin{subequations}
        \begin{align}
            \rho \frac{\partial \bs{u}}{\partial t} + \rho (\bs{u} \cdot \nabla)\bs{u} -
            \mu \Delta \bs{u} + \nabla p = 0 \text{ in } \Omega \\
            \nabla \cdot \bs{u} = 0 \text{ in } \Omega
            \\
            \bs{u} = \bs{u}_{in} \text{ on } \Gamma_{in}
            \\
            \bs{u} = \bs{0} \text{ on } \Gamma_w
            \\
            \mu \frac{\partial \bs{u}}{\partial \bs{n}} - p\bs{n} = -P_l(t)\bs{n} \text{
            on } \Gamma_l
            \label{eq:forward}
        \end{align}
    \end{subequations}
    with $\rho, \mu$ the density and dynamic viscosity of the fluid and $P_l(t)$ being
    given by a Windkessel boundary condition defined by:
    \begin{subequations}
        \begin{align}
            P_l = R_{p,l}Q_l + \pi_l                  \\
            Q_l = \int_{\Gamma_l} \bs{u} \cdot \bs{n} \\
            C_{d, l} \frac{d\pi_l}{dt} + \frac{\pi_l}{R_{d, l}} = Q_l
        \end{align}
    \end{subequations}
    This boundary condition models the effects of the remaining vascular system on the
    outlet via the proximal and distal resistances $R_p$, $R_d$ of the vasculature and the
    distal compliance $C_d$ of the vessels.
    Lastly, the inflow $\bs{u}_{in}$ is defined as
    \begin{equation*}
        \bs{u}_{in} = -U f(t) \bs{n}  \,,
    \end{equation*}
    where $U$ is a constant amplitude and
    \begin{equation*}
        f(t) =
        \begin{cases}
            \sin(\frac{\pi t}{T}) & \text{if } t \leq T \,,\\
            \frac{\pi}{T}(t-T)\exp^{-k(t-T)} & \text{if } T_c > t > T \,,
        \end{cases}
    \end{equation*}
    with $T_c = 0.8$ and $T = 0.36$.
    The physical parameters are set as seen in \Cref{tab:Parameters}.
    \begin{table}[!hbtp]
        \centering
        \begin{tabular}{|c|c|}
            \hline
            Parameter                & Value   \\
            \hline
            $\rho \ (gr \cdot cm^3)$ & $1.2$   \\
            $\mu \ (P)$              & $0.035$ \\
            $U \ (cm \cdot s^{-1})$  & $75$    \\
            $T_c \ (s)$              & $0.80$  \\
            $T \ (s)$                & $0.36$  \\
            $\kappa \ (s^{-1})$      & $70$    \\
            \hline
        \end{tabular}
        \quad
        \begin{tabular}{|c|c|c|c|c|}
            \hline
            & $ \Gamma_{1}$    & $\Gamma_{2}$     & $\Gamma_{3}$     & $\Gamma_{4}$     \\
            \hline
            $R_p \ (dyn \cdot s \cdot cm^{-5})$ & $480$            & $520$            &
            $520$            & $200$            \\
            $R_d \ (dyn \cdot s \cdot cm^{-5})$ & $7200$           & $11520$          &
            $11520$          & $4800$           \\
            $C \ (dyn^{-1} \cdot cm^5 )$        & $4\cdot 10^{-4}$ & $3\cdot 10^{-4}$ &
            $3\cdot 10^{-4}$ & $4\cdot 10^{-4}$ \\
            \hline
        \end{tabular}
        \caption{Physical parameters and numerical values of the three-element Windkessel
            parameters for every outlet.
        } \label{tab:Parameters}
    \end{table}
    The forward problem is solved using an in-house finite elements solver, with a
    semi-implicit 3D-0D coupling scheme as in  \cite{garay_parameter_2022}, using $P1$
    elements for both the velocity and the pressure and a time step of $dt = 1\,ms$. The
    full algorithm is detailed in the appendix \Cref{sec:app_NS}.

    \subsection{Measurement problem}

    The forward solution is undersampled in time to $dt_{meas} = 15\,ms$, resulting in a
    total of 56 measurements of the velocity. From the solution of the forward
    problem, we simulate a
    PC-MRI acquisition by sub-sampling onto a rectangular domain meshed in a structured
    fashion with a resolution of $2\,mm$ in all three spatial directions, and then
    applying the process described in Equation \eqref{eq:freq_meas} with a $venc$ of twice
    the maximal velocity. The magnitude is modeled as
    \begin{equation}
        M(\bs{x}) =
        \begin{cases}
            1.0 & \text{if } \bs{x} \text{ is in the lumen of the vessel}\,, \\
            0.5 & \text{otherwise} \,.
        \end{cases}
    \end{equation}
    and the background phase was set to an arbitrary constant value of $\phi_{back} =
    7.5\cdot 10^{-2}rad$.

    The unknown parameters are taken to be the inflow $U$ and the distal resistances
    $R_{d, i}$ of the Windkessel outlets $\Gamma_1, \Gamma_2, \Gamma, 3$. We have
    decided to optimize only a two-dimensional mask in the $x-y$-plane and
    ``stack" these masks for the $z$-dimension, which is a common approach for 3D MRI
    acquisitions. We are also using the same masks for all velocity directions.
    Finally a complex Gaussian noise $\bs{\epsilon} \in \mathbb{C}^N$ is added with a
    signal-to-noise ratio (SNR) of 15. Fifty independent realizations of the noise
    were generated.

    \subsection{Sensitivities}

\begin{figure}[htbp!]
        \centering
        \subfloat[$U$]{
        \includegraphics[height=0.2\textheight]{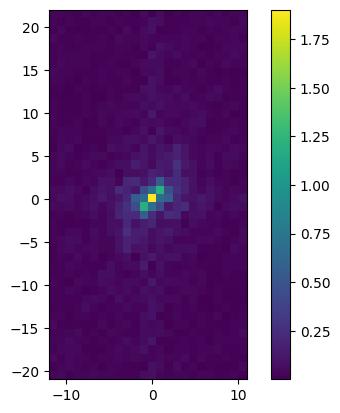}}
        \subfloat[$R_{d, 1}$]{
        \includegraphics[height=0.2\textheight]{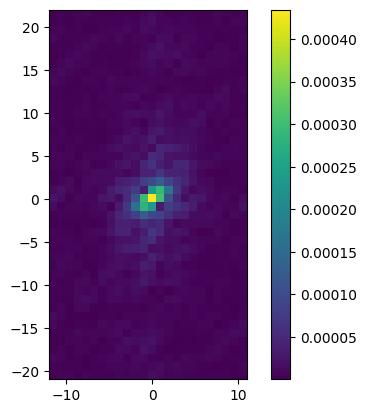}}
        \subfloat[$R_{d, 2}$]{
        \includegraphics[height=0.205\textheight]{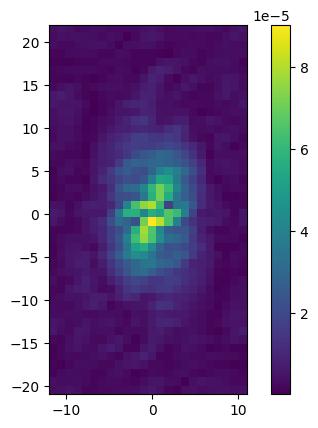}}
        \subfloat[$R_{d, 3}$]{
        \includegraphics[height=0.205\textheight]{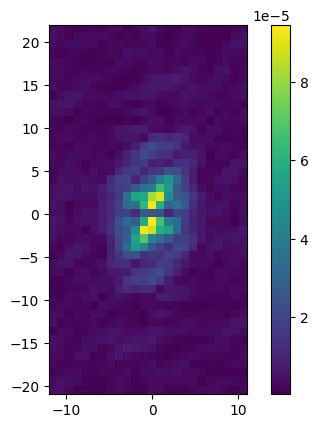}}
        \caption{Magnitudes of the sensitivities of the measurement to each of the
        parameters in the aorta test case. Note the different scales shown in the color bars.}
        \label{fig:aorta_sens}
    \end{figure}

    The sensitivities for each parameter are computed numerically using a first-order
    difference scheme with a relative difference of $h = 10^{-6}$.
    The magnitude of the sensitivities for each parameter for a slice at $z = 0$ is shown
    in \Cref{fig:aorta_sens}. It is already evident that the sensitivities of the
    parameters follow different patterns, with those of the inflow $U$ especially focused
    on the center while those of the Windkessel resistances spread out more. Note also
    that the sensitivity of $U$ is several orders of magnitude larger than the other
    sensitivities. Due to this, as well as the inflow usually being much more robust in
    estimation than the Windkessel parameters, we have chosen to use only the
    sensitivities of the three Windkessel resistances for the construction of the optimal masks.

    \subsection{Implementation of the optimization algorithm}

    We create masks for the same four budgets of points $N_s = 5, 10, 25 and 50$
    (in the 2D mask, hence more points in the complete 3D mask) which now corresponds
    to acceleration
    factors $R \sim 211, 106, 42, 21$, respectively.

    We are using the same implementation as for the analytic test case. The compute
    times are listed in \Cref{tab:timing_aorta}. With the higher computational cost
    per computation of the FIM, the advantage of the greedy algorithm is now evident for
    all budgets except the lowest one. Especially for the highest budget, the exchange
    algorithm requires several hours to complete for all three parameters, whereas the
    greedy algorithm finishes after a few minutes.

    \begin{table}[htbp!]
        \centering
        \begin{tabular}{|c|c|c|c|c|c|c|c|c|}
            \hline
            Algorithm & \multicolumn{4}{c}{1 parameter} & \multicolumn{4}{|c|}{3
            parameters}
            \\ \cline{2-9}
            & 5                               & 10                                 & 25
            & 50      & 5     & 10       & 25       & 50       \\
            \hline
            Greedy    & 8.92s                           & 28.93s
            & 1min53s & 3min56s & 9.32s & 19.78s   & 1min07s  & 4min 34s \\
            Exchange  & 1.1s                            & 27.38s
            & 1min28s & 1h07min & 10s   & 1min 38s & 27min27s & 2h18min  \\
            \hline
        \end{tabular}
        \caption{Computing times for the two algorithms for different budgets on the
        aorta testcase}
        \label{tab:timing_aorta}
    \end{table}

    \subsection{Optimal masks}

\begin{figure}[htbp!]
        \centering
        \subfloat[A-crit., greedy]{
        \includegraphics[width=0.18\textwidth]{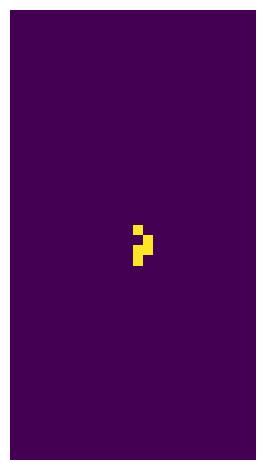}}\qquad
        \subfloat[A-crit., exchange]{
            \includegraphics[width=0.18\textwidth]{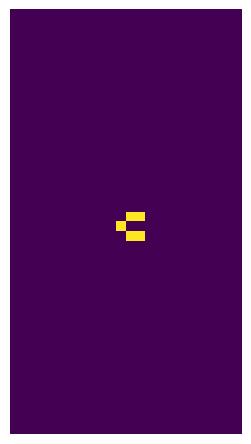}
        }\qquad
        \subfloat[D-crit., greedy]{
            \includegraphics[width=0.18\textwidth]{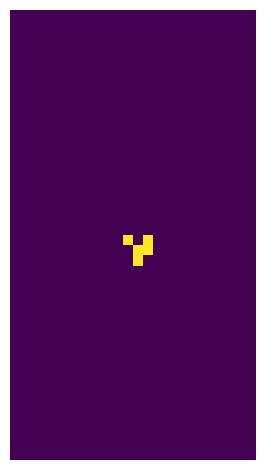}
        }\qquad
        \subfloat[D-crit., exchange]{
            \includegraphics[width=0.18\textwidth]{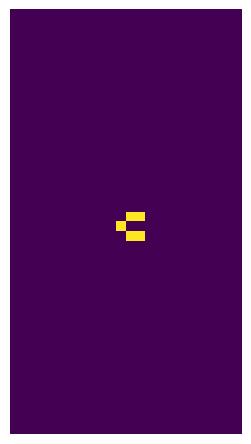}
        }
        \caption{Masks for a budget of $N_S = 5$ frequencies.
        }
        \label{fig:aorta_budget_5}
    \end{figure}

\begin{figure}[htbp!]
        \centering
        \subfloat[A-crit., greedy]{
        \includegraphics[width=0.18\textwidth]{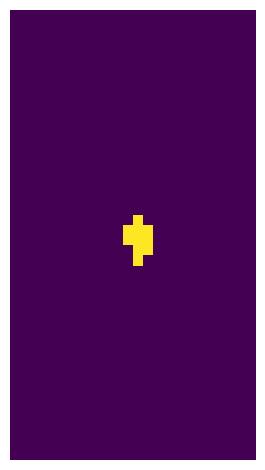}}\qquad
        \subfloat[A-crit., exchange]{
            \includegraphics[width=0.18\textwidth]{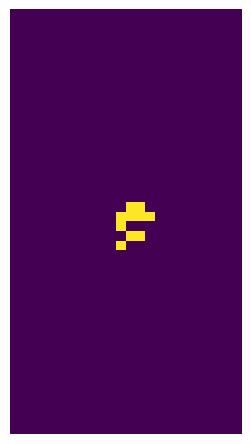}
        }\qquad
        \subfloat[D-crit., greedy]{
            \includegraphics[width=0.18\textwidth]{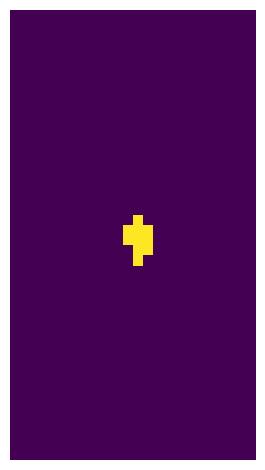}
        }\qquad
        \subfloat[D-crit., exchange]{
            \includegraphics[width=0.18\textwidth]{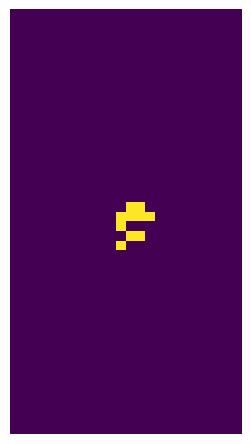}
        }
        \caption{Masks for a budget of $N_S = 10$ frequencies.}
        \label{fig:aorta_budget_10}
    \end{figure}

\begin{figure}[htbp!]
        \centering
        \subfloat[A-crit., greedy]{
        \includegraphics[width=0.18\textwidth]{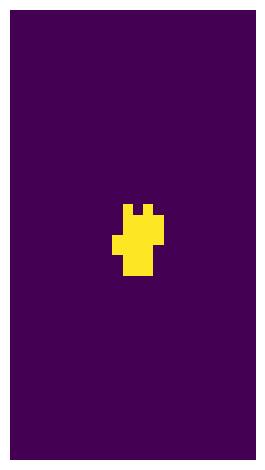}}\qquad
        \subfloat[A-crit., exchange]{
            \includegraphics[width=0.18\textwidth]{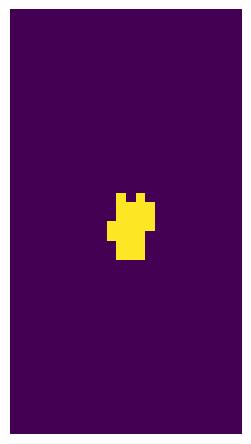}
        }\qquad
        \subfloat[D-crit., greedy]{
            \includegraphics[width=0.18\textwidth]{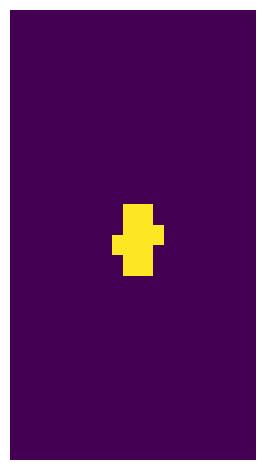}
        }\qquad
        \subfloat[D-crit., exchange]{
            \includegraphics[width=0.18\textwidth]{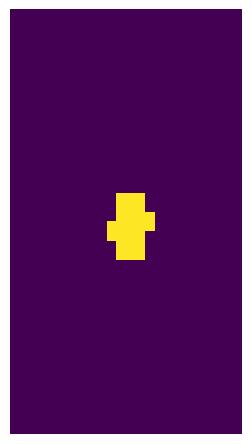}
        }
        \caption{Masks for a budget of $N_S = 25$ frequencies.}
        \label{fig:aorta_budget_25}
    \end{figure}

\begin{figure}[htbp!]
        \centering
        \subfloat[A-crit., greedy]{
        \includegraphics[width=0.18\textwidth]{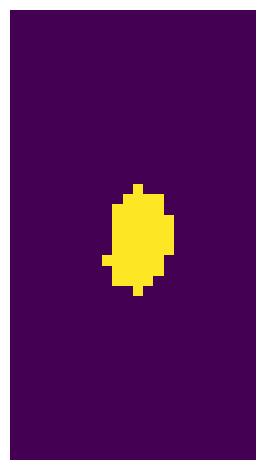}}\qquad
        \subfloat[A-crit., exchange]{
            \includegraphics[width=0.18\textwidth]{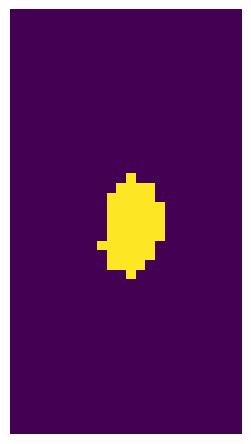}
        }\qquad
        \subfloat[D-crit., greedy]{
            \includegraphics[width=0.18\textwidth]{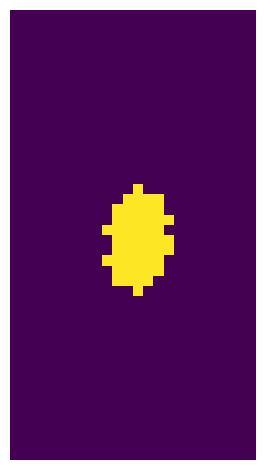}
        }\qquad
        \subfloat[D-crit., exchange]{
            \includegraphics[width=0.18\textwidth]{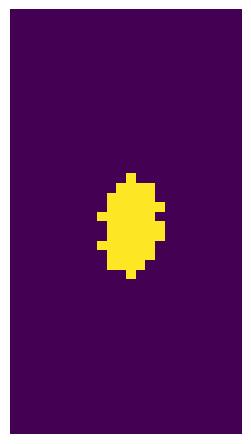}
        }
        \caption{Masks for a budget of $N_S = 50$ frequencies.}
        \label{fig:aorta_budget_50}
    \end{figure}

    The optimal masks corresponding to the selected budgets are shown in Figures
    \ref{fig:aorta_budget_5}---\ref{fig:aorta_budget_50}.  Overall, the optimal masks have
    an tilted, elliptic shape, populating the center of the k-space. For the budgets of 5
    and 10 points, considerable differences between the results of the different
    algorithms can be seen. For the higher budgets, all generated masks are
    near-identical, and even for the lower budgets, the A- and D-optimal masks are
    often identical.

    \subsection{Conventional masks}

\begin{figure}[htbp!]
        \centering
        \subfloat[Gaussian, budget 5]{
        \includegraphics[width=0.25\textwidth]{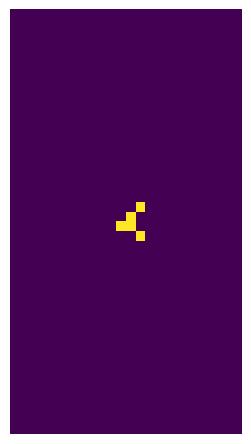}}
        \subfloat[Gaussian, budget 10]{
        \includegraphics[width=0.25\textwidth]{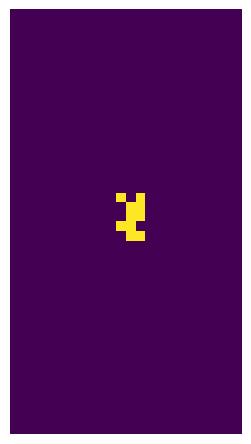}}
        \subfloat[Gaussian, budget 25]{
        \includegraphics[width=0.25\textwidth]{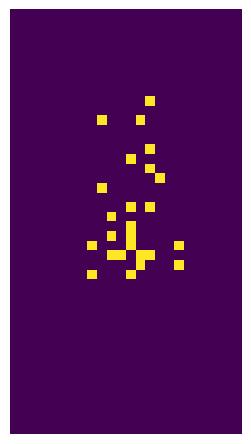}}
        \subfloat[Gaussian, budget 50]{
        \includegraphics[width=0.25\textwidth]{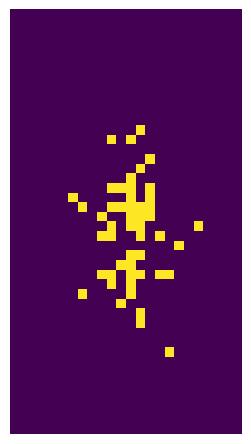}}\\
        \subfloat[circle, budget 5]{
        \includegraphics[width=0.25\textwidth]{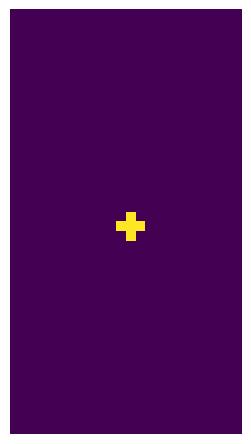}}
        \subfloat[circle, budget 10 (9 points)]{
        \includegraphics[width=0.25\textwidth]{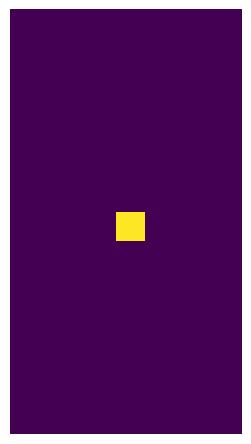}}
        \subfloat[circle, budget 25]{
        \includegraphics[width=0.25\textwidth]{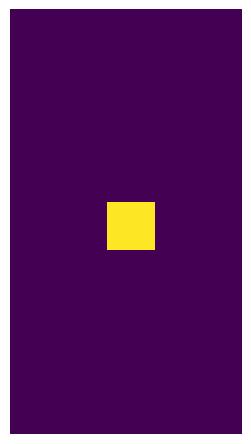}}
        \subfloat[circle, budget 50 (49 points)]{
        \includegraphics[width=0.25\textwidth]{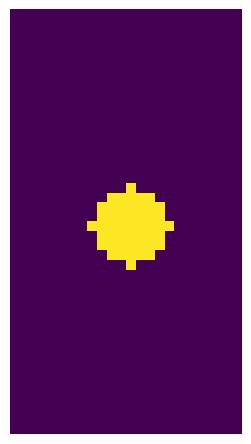}}
        \caption{Masks generated using a pseudo-random Gaussian sampling pattern and a
            circular sampling pattern, for a variety of acceleration factors, for the
        aorta test case.}
        \label{fig:aorta_conv_masks}
    \end{figure}

    We are again using the ``circle" and the pseudo-random Gaussian pattern for
    comparison, which are shown in \Cref{fig:aorta_conv_masks}.

    \subsection{Parameter estimation}

    In this case we generate fifty independent noise realizations with an SNR of 15 and
    use ten iterations of the ROUKF, each time using the result of the last iteration
    as the new initial guess of the next, but resetting the standard deviation of the
    parameters. The true values are $R_{d,1} = 7200, R_{d, 2} =
    11520, R_{d, 3} = 11520$, while the initial values provided to ROUKF were $R_{d, 1} =
    4000, R_{d, 2} = 4000, R_{d, 3} = 4000$.
    \begin{figure}[htbp!]
        \centering
        \subfloat[A-opt., exchange]{
            \includegraphics[width=0.5\textwidth]{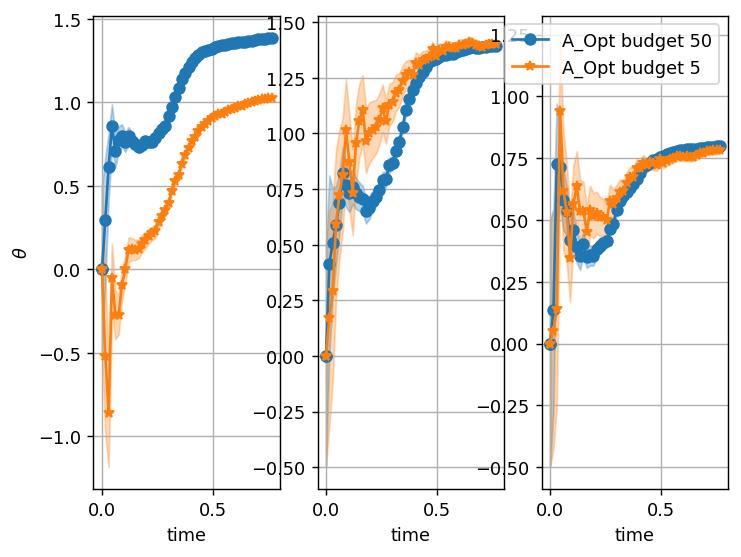}
        }
        \subfloat[circle]{
            \includegraphics[width=0.5\textwidth]{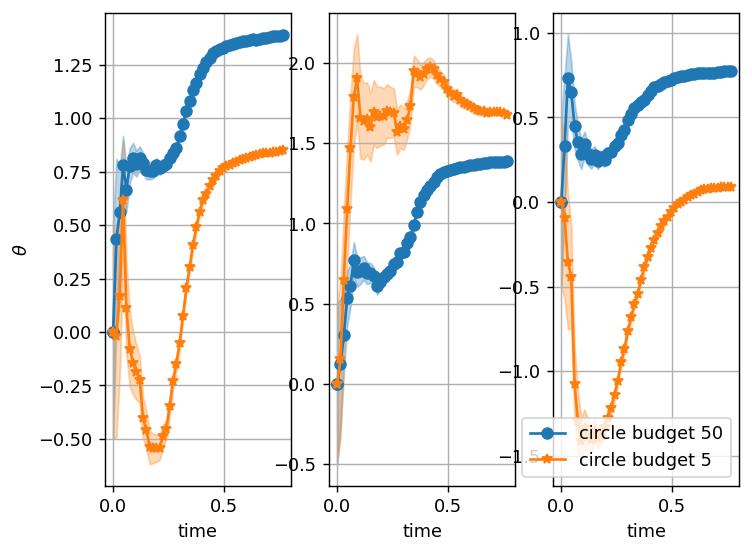}
        }
        \caption{Evolution of the parameters over time during the first iteration of
            the ROUKF, for
        two different budgets.}
        \label{fig:param_evolution}
    \end{figure}

    We show the parameter evolution curves of the ROUKF algorithm for the circle
    mask and the A-optimal mask with the exchange algorithm in
    \Cref{fig:param_evolution} for the first ROUKF iteration. It can be seen
    that the curves for a lower budget contain more and higher spikes in either direction
    before converging, for either mask. While both curves qualitatively follow the same
    shape, it can be seen that the curves for the circle mask do not reach the same
    values, indicating a difference in assimilated information. Meanwhile, the curves for
    the optimal masks remain much closer throughout the entire time. This seems to
    indicate a smaller loss of information due to still selecting the most
    information-dense points.

\begin{figure}[htbp!]
        \centering
        \includegraphics[width=0.6\textwidth]{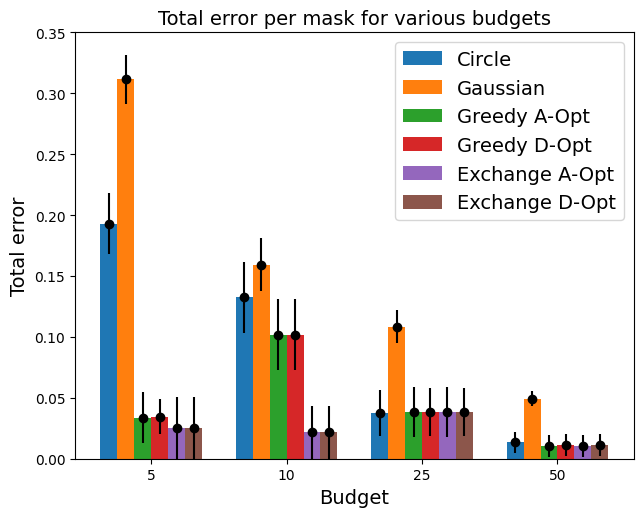}

        \caption{Total error of the Windkessel parameters estimated in the inverse problem for
            the A-optimal and D-optimal masks, as well as the circle and Gaussian masks, for
        various budgets.}
        \label{fig:aorta_errors}
    \end{figure}

    The total error $e = \frac{||\bs{\theta} -
    \bs{\theta}_{true}||}{||\bs{\theta}_{true}||}$ of the three estimated parameters,
    for all four budgets and the
    circle, Gaussian, and all different optimal masks, are shown in
    \Cref{fig:aorta_errors}. For all budgets, but especially for the lower budgets, the
    optimal masks considerably outperforms the conventional masks in terms of error. The
    variance of the error of the parameter estimates remains similar between the different masks.

    Comparing the different algorithms, there is no difference for the higher budgets (25
    and 50) as here both algorithms led to the same mask. For the lower budgets, however,
    there is a notable advantage in using the exchange algorithm over the greedy algorithm.
    There are only minimal differences in the results between the A- and D- optimal masks.
    The exception is for a budget of 5 points, where a single point being in a different
    position in the masks identified by the greedy algorithm does lead to a higher error
    for the A-optimal mask.

    We can observe the error standard
    deviation consistently decreases when increasing the budget, as expected. The error
    mean also decreases, with a single exception of the exchange algorithm when increasing
    the budget from 10 to 25. Though we have no clear explanation about this result, a
    possible explanation could be that the additional selected frequencies lead to a cost
    function with a local minimum that is closer to the initial guess, hence making the
    ROUKF algorithm converge to such result.

\begin{figure}
        \centering
        \subfloat[Parameter 1]{
            \includegraphics[width=0.5\textwidth]{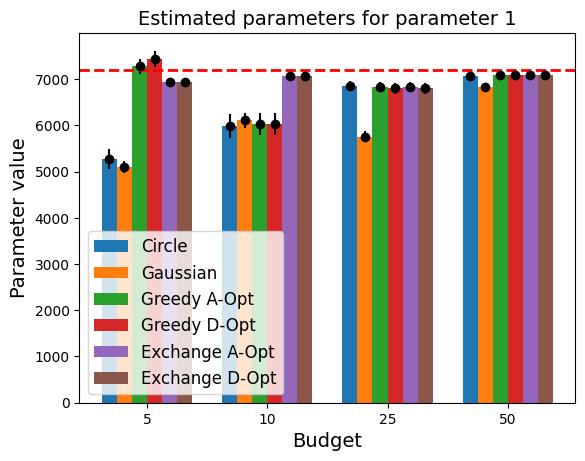}
        }
        \subfloat[Parameter 2]{
            \includegraphics[width=0.5\textwidth]{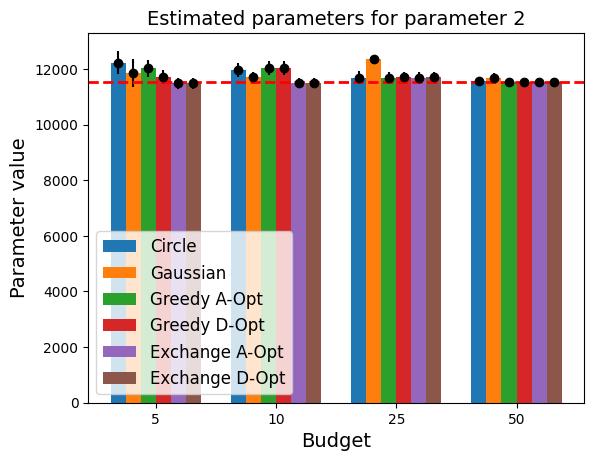}
        }\\
        \subfloat[Parameter 3]{
            \includegraphics[width=0.5\textwidth]{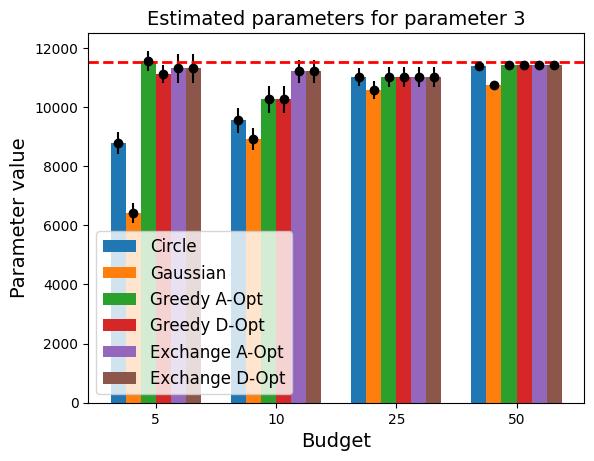}
        }
        \caption{Estimated values of the Windkessel parameters for different masks. Red dashed
        lines indicate true values of the parameters.}
        \label{fig:aorta_param_bars}
    \end{figure}

    We display the estimated values of each parameter for the different masks and
    budgets in \Cref{fig:aorta_param_bars}. Both qualitatively and quantitatively,
    these results conform with the observations above. The difference in performance
    between the optimal masks and the conventional masks is more pronounced for $R_{d,1}$
    and $R_{d, 3}$ than $R_{d, 2}$, though, which could indicate an advantage of the
    optimal masks especially in estimating difficult parameters (as those parameters
        correspond to smaller outlets compared to $R_{d,2}$, and are therefore in our
    experience harder to estimate accurately).

\begin{figure}[htbp!]
        \centering
        \subfloat[A-optimality criterion]{
            \includegraphics[width=0.5\textwidth]{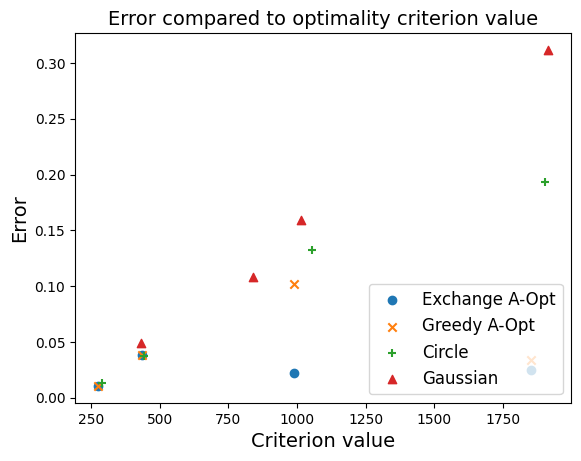}
        }
        \subfloat[D-optimality criterion]{
            \includegraphics[width=0.5\textwidth]{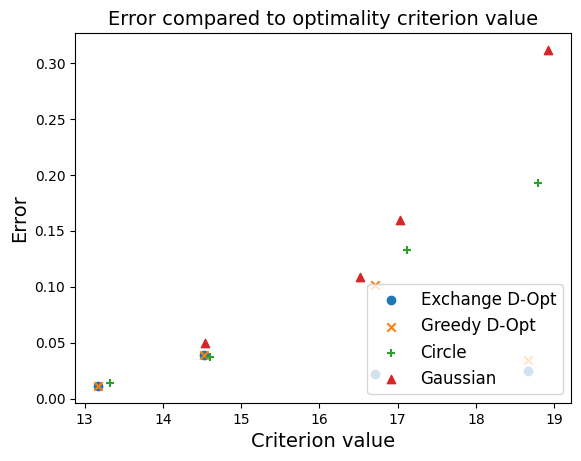}
        }

        \caption{Optimality criterion values compared to the errors of the estimated
            parameters in the aorta test case. A lower optimality criterion value is better. Due
        to overlaps, not all points may be clearly visible.}
        \label{fig:aorta_crit_values}
    \end{figure}

\begin{figure}
        \centering
        \subfloat[A-optimality criterion]{
            \includegraphics[width=0.5\textwidth]{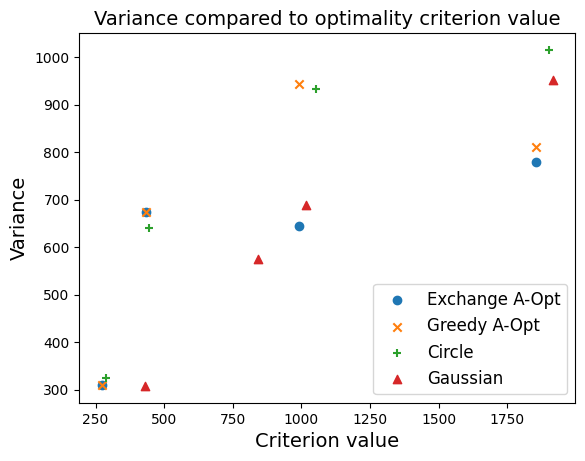}
        }
        \subfloat[D-optimality criterion]{
            \includegraphics[width=0.5\textwidth]{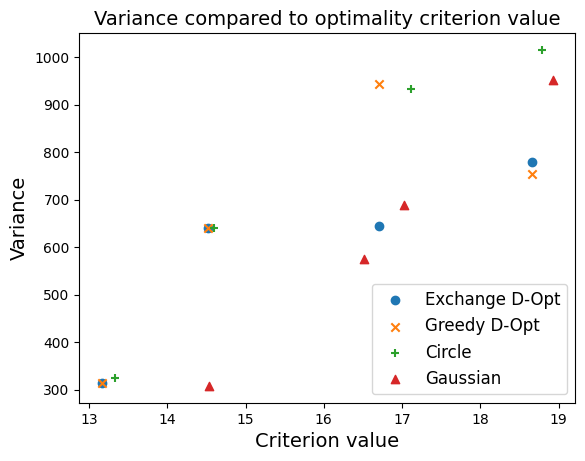}
        }

        \caption{Optimality criterion values compared to the sum of the variances of the
            estimated parameters of the aorta test case. A lower optimality criterion value is
        better. Due to overlaps, not all points may be clearly visible.}
        \label{fig:aorta_crit_vars}
    \end{figure}

    We also plot the error and the sum of the variances against the
    criterion value in Figures \ref{fig:aorta_crit_values}, and \ref{fig:aorta_crit_vars},
    respectively. The visible trend confirms the relation between the criteria value and
    the outcome of the inverse problem also in this more complex test case.

    \subsection{Using sensitivities for estimated parameter values}

    Here as well we consider sensitivities computes with incorrect parameter values, in
    this case, the same ones as the ones used as the initial guess for the inverse problem.

\begin{figure}[htbp!]
        \centering
        \subfloat[exchange algorithm]{
        \includegraphics[width = 0.5\textwidth]{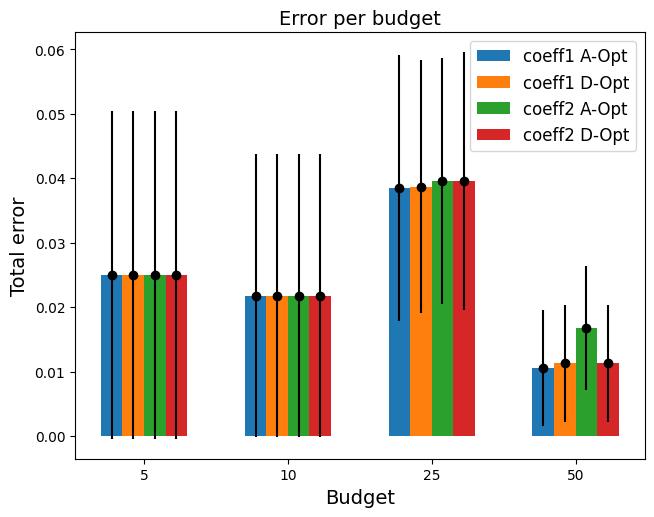}}
        \subfloat[greedy algorithm]{
        \includegraphics[width = 0.5\textwidth]{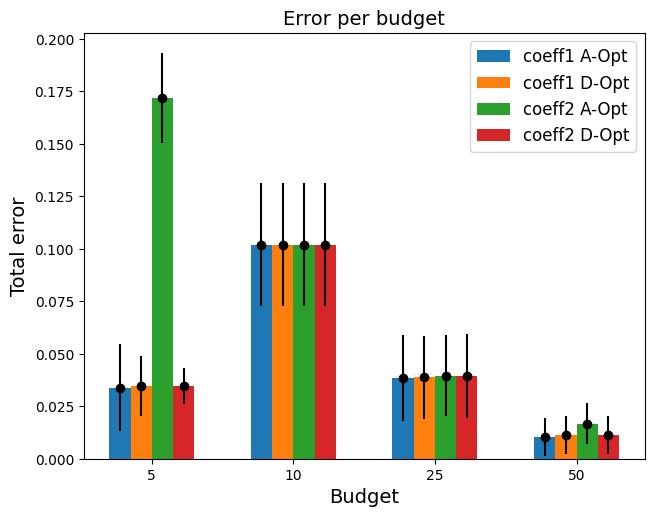}}
        \caption{Comparison of the error of the inverse problem between the sensitivities
        computed with the true parameter values and the incorrect parameter values.}
        \label{fig:aorta_approx_sens}
    \end{figure}

    The optimal masks computed with these sensitivities, like for the analytical test
    case \eqref{eqn:analytical_fun}, often agree with those computed
    with the true sensitivities. The results of the
    inverse problem using the masks, for the true and approximated sensitivities, are
    shown in \Cref{fig:aorta_approx_sens}. It can be seen that using the
    approximated sensitivities does not lead to significantly different results. The
    process therefore seems robust to the exact parameter values used for the computation
    of the sensitivities, at least within a realistic range.

\begin{figure}[htbp!]
        \centering
        \subfloat[accurate sensitivities]{
            \includegraphics[width=0.25\textwidth]{figures/greedy/p1-3/mask_A_5_aorta_greedy.jpg}
        }
        \subfloat[approximate sensitivities]{
            \includegraphics[width=0.25\textwidth]{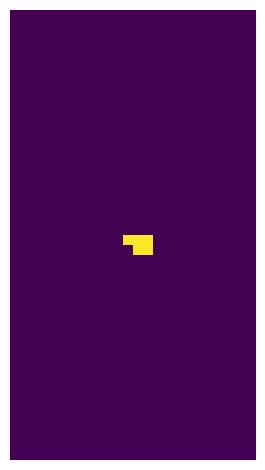}
        }
        \caption{A-optimal masks optimized with the greedy algorithm, for sensitivities
        computed with the true parameter values and with incorrect parameter values.}
        \label{fig:approx_comp}
    \end{figure}

    The exception is the A-optimal greedy mask for the lowest budget, which shows a
    considerably higher error when using sensitivities generated with approximated values.
    The masks for this case, as well as the corresponding masks with the accurate
    sensitivities, are shown in \Cref{fig:approx_comp}, and are one of the few sets
    of masks that differ notably based on the change in the sensitivities. Therefore for
    very low budgets, with approximated sensitivities, the exchange algorithm may be a
    more reliable choice. Nevertheless, even in this case the error for the greedy mask is
    less than for a conventional mask.

    \section{Discussion}\label{sec:discussion}

    The numerical results are generally consistent between the analytical test case and
    the aorta test case. The optimal masks outperform the conventional masks in terms of
    both error and variance in nearly every case, for both optimality criteria and for
    either optimization algorithm. 

    The results for the aorta test case are more similar between criteria and algorithms,
    which may be due to the different nature of the underlying signal. As a result, no
    clear preference between the optimality criteria is apparent. Other optimality
    criteria could be explored, which may have a better relation to the error of the
    estimated parameters. A criterion such as the c-criterion, which optimizes based
    on a predetermined linear combination of the parameters, could also be used to
    account for differences in the scale of the sensitivities for different
    parameters\cite{alexanderian_optimal_2021}.

    Other optimization methods should be explored as well, since the exchange algorithm
    may be unfeasible for an increased number of parameters or higher data resolution, and
    the greedy algorithm may not lead to an optimal solution, especially for more complex
    problems. A potential solution could be to initialize the exchange algorithm with the
    results of the greedy algorithm, and therefore potentially cut down on the number of
    exchanges required compared to a randomly initialized mask, or to use a more
    optimized version of the algorithm. Another option may be to
    use probabilistic/stochastic algorithms \cite{attia2022stochastic,attia2024probabilistic}
    to find the optimal design by iteratively sampling from the
    distribution. Alternatively, if this is computationally unfeasible, we could relax the
    constraint of having a discrete design and instead consider a function of weights in
    $[0, 1]^N$ such that they sum up to 1, leading to a continuous optimization problem.
    This raises additional questions about how to round the resulting continuous design to
    obtain a discrete design, and may also not result in the optimal discrete design;
    see e.g., \cite{attia2022optimal}.

    An alternative approach would be to work with a parametrizable family of masks, with
    additional constraints to ensure they result in a valid sampling pattern
    \cite{ermakov_mathematical_1983}. However,
    this would only provide an optimal design within this family of masks.

    Additionally, a further exploration of the effects of the assumed parameters for the
    computation of the sensitivities is needed, as it is likely that the computed design
    holds only within a range of coefficients which are close enough. This uncertainty in
    the coefficients and therefore in the sensitivities can also be included in the
    optimal design process, using techniques such as in
    \cite{fedorov_model-oriented_2025, alexanderian_optimal_2025}.

    One limitation of the numerical analysis carried out in this paper is the lack of real data.
    Real MRI data generally
    includes measurements taken by different magnet coils. Due to their different positioning,
    these coils generally have a different ability to detect the signal in the field-of-view,
    which is often non-uniform in space as well. This is often shown in the form of a "coil
    sensitivity map", which describes how well a particular coil captures the signal
    in a specific
    voxel. The effect of taking measurements with these differing coil sensitivities will likely
    have to be accounted for. Additionally, the magnitudes and background phases can vary in time
    and space as well, which may need to be considered in the computation of the measurement
    sensitivities for the optimal design process. As our method requires
    the magnitude and the background phase, it will have to be explored whether the
    measurement sensitivities computed with estimations or modelled versions result in
    sufficiently
    good optimal designs. We plan to address this in future research.

    \section{Conclusion}\label{sec:conclusion}

    In this work, we have introduced an Optimal Experimental Design (OED) framework for
    the selection of k-space sampling patterns tailored to parameter estimation in inverse
    hemodynamics problems. Through both analytical and aortic hemodynamics test cases, we
    demonstrated that masks optimized using OED criteria (A- and D-optimality)
    consistently outperform conventional sampling patterns in terms of parameter
    estimation accuracy and variance, especially for high undersampling (acceleration
    factor $R \approx 200$). The optimal masks achieve lower errors with a
    budget of 5 points per slice than the conventional masks with a budget of $50$, thus
    providing a speed-up of over $10\times$ in terms of the acquisition time in the flow encoding
    directions. The results also show that the choice of optimization algorithm can impact
    performance for low sampling budgets, with the exchange algorithm providing some
    advantage over the greedy approach in these cases, but at a higher computational cost.

    Our findings indicate that OED-based mask design is a promising strategy for
    accelerating MRI acquisitions while maintaining or improving the quality of parameter
    estimates in cardiovascular modeling. Though further research is needed, the approach
    is robust to moderate inaccuracies in the assumed parameter values used for
    sensitivity calculations, and the optimality criteria correlate well with actual
    estimation errors. Future work should address the integration of more advanced
    optimization algorithms, the extension to real MRI data with coil sensitivities, and
    the incorporation of parameter uncertainty into the design process to further enhance
    the practical applicability of this framework.

    \section*{Acknowledgments}
    C.B. and M.L. acknowledge the funding from the European Research Council (ERC)
    under the European Union's Horizon 2020 research and innovation program (grant
    agreement No 852544 - CardioZoom). A. A. was supported by the Applied Mathematics
    activity within the U.S. Department of Energy, Office of Science, Advanced
    Scientific Computing Research, under Contract DEAC02-06CH11357.

    \bibliography{bib.bib}

    \appendix
    \section{Numerical solution method of the flow model}\label{sec:app_NS}
    Here we detail the algorithm used to solve the incompressible Navier-Stokes
    equation with Windkessel boundary conditions for the forward problem.

    \begin{algorithm}[H]
        \caption{Fractional step algorithm with a modified semi-implicit Windkessel
        model coupling}
        \label{alg:semi}
        Given the initial conditions $\bs{u}^0 = \bs u(0) \in V_{\Gamma_{w},h}$ and
        $\pi^0_1,\dots,\pi_N^0 \in \mathbb{R}$,  perform for $j>0$, with $t^{j}  = j \tau$: \\

        \textbf{1. Viscous Step:} Find the tentative velocity $\tilde{\bs{u}}^{n} \in
        V_{\Gamma_{w},h}$ such that:
        \begin{equation}
            \begin{cases}
                \displaystyle \tilde{\bs{u}}^{j}|_{\Gamma_{in}} = \bs{u}_{inlet}(t^{j}) \\
                \displaystyle \frac{\tau}{\rho} (\tilde{\bs{u}}^{j} , \bs{v}
                )_{\Omega_h} + \rho ( \bs{u}^{j-1} \cdot \nabla \tilde{\bs{u}}^{j} ,
                \bs{v}  )_{\Omega_h} + \frac{\rho}{2} ( (\nabla \cdot \bs{u}^{j-1})
                \tilde{\bs{u}}^{j}, \bs{v}  )_{\Omega_h}
                + (\delta \bs{u}^{j-1} \cdot \nabla \tilde{\bs{u}}^j , \bs{u}^{j-1}
                \cdot \nabla \bs{v}  )_{\Omega_h} \\
                \displaystyle + 2\mu (\epsilon(\tilde{\bs{u}}^{j}) ,
                \epsilon(\bs{v}))_{\Omega_h} + \sum_{\ell=1}^K \frac{\rho}{2} |
                \bs{u}^{j-1}\cdot \mathbf{n} |_{-}
                (\tilde{\bs{u}}^j,\bs{v})_{\Gamma_\ell} = \frac{\tau}{\rho}
                (\bs{u}^{j-1} , \bs{v})_{\Omega_h}
            \end{cases}
        \end{equation} \label{eq:viscousstep}
        for all $\mathbf{v} \in V_{\Gamma_{in}\cup \Gamma_{w} ,h}$, and $|x|_{-}$
        denoting the negative part of $x$.

        \textbf{2. Projection-Windkessel Step:} Compute $\tilde{Q}^{j} =
        \int_{\Gamma_\ell} \tilde{\bs{u}}^{j} \cdot \bs{j}$. Find $p^{j} \in Q_h$ such that:
        \begin{equation} \label{eq:pressproj}
            \displaystyle \frac{\tau}{\rho} (\nabla p^{j} , \nabla q)_{\Omega_h} +
            \sum_{\ell=1}^K \frac{ \overline{p^{j}}_{\Gamma_\ell}
            \ \overline{q}_{\Gamma_\ell}}{\gamma_\ell} +  \epsilon
            \sum_{\ell=1}^K(\mathcal{T}(\nabla p^j), \mathcal{T}(\nabla
            q))_{\Gamma_\ell} =  \sum_{\ell=1}^K \bigg( \tilde{Q}^{j} +
            \frac{\alpha_\ell \pi_\ell^{j-1}}{\gamma_\ell}  \bigg)
            \overline{q}_{\Gamma_\ell} - (\nabla \cdot \tilde{\bs{u}}^{j} , q)_{\Omega_h} ,
        \end{equation}
        for all $q \in Q_h$ and with $\overline{(\cdot)}_{\Gamma_{\ell}} =
        \frac{1}{Area(\Gamma_\ell)} \int_{\Gamma_\ell} (\cdot) ds$ and
        $\mathcal{T}(\mathbf{f}) = \mathbf{f} - (\mathbf{f}\cdot\mathbf{j}) \mathbf{j}$.

        \textbf{3. Velocity correction Step:} Find $\bs{u}^{j} \in [L^2(\Omega_h)]^3 $ such that:
        $$\displaystyle (\bs{u}^{j},\mathbf{v})_{\Omega_h} = (\tilde{\bs{u}}^{j} -
        \frac{\tau}{\rho} \nabla p^{j},\mathbf{v})_{\Omega_h}
        $$
        for all $\mathbf{v} \in [L^2(\Omega_h)]^3$

        \textbf{4. Update-Windkessel Step:} Set $P^{j}_\ell =
        \overline{p^{j}}_{\gamma_\ell} $ and compute $\pi_\ell^{j} \in \mathbb{R}$ as:
        $$ \pi_\ell^{j} = \big ( \alpha_\ell - \frac{\alpha_\ell
        \beta_\ell}{\gamma_\ell} \big ) \ \pi_\ell^{j-1} +
        \frac{\beta_\ell}{\gamma_\ell} \  P^{j}_\ell \ , \ \ell = 1,...,K $$
    \end{algorithm}
    \section{The Reduced-Order Unscented Kalman Filter}
    \label{sec:app_ROUKF}
    The ROUKF algorithm is as follows:

    Let us first consider the notation $\bs{Z}_{(*)}$ as the matrix with the column-wise
    collection of vectors $\bs{Z}_{(1)},\bs{Z}_{(2)},\dots$.

    Define  the  \textit{canonical sigma-points}  $\bs{I}_{(1)},\dots,\bs{I}_{(2p)} \in
    \mathbb{R}^p$ such that
    \begin{equation}
        \bs{I}_{(j)} =
        \begin{cases}
            \sqrt{p}\bs{e}_j,      & \text{for } 1\leq j \leq p     \\
            -\sqrt{p}\bs{e}_{j-p}, & \text{for } p+1 \leq j \leq 2p
        \end{cases}
    \end{equation}
    where the vectors $\bs{e}_i$ form the canonical base of of $\mathbb{R}^p$.
    Moreover, define the weight $\alpha = \frac{1}{2p}$.

    We denote by $\bs{\hat X}^n_-,\bs{\hat X}^n_+ \in\mathbb{R}^r$ a priori (model
    prediction) and a
    posteriori (corrected by observations) estimates of the true state
    $\bs{X}^n\in\mathbb{R}^r$. In our PDEs, this consists of the velocity field $\bs{X}^n
    = \mathcal{A}^n(\bs{\theta})$.

    For given values of the initial condition $\bs{\hat{X}}^0_+ = \bs{X}_0 \in \mathbb{R}^r$,
    the initial expected value of the parameters $\bs{\hat{\theta}}^0_+ =
    \bs{\theta}_0 \in\mathbb{R}^p$ and its covariance matrix $\bs{P}_0$, perform
    \begin{itemize}
            \begin{subequations}\label{eq:ParamSEIK}
            \item \textbf{Initialization:}  initialize the sensitivities as
                \begin{multline}
                    \shoveright{\hspace{0.1cm}
                        \bs{L}_{\theta}^{0} = \sqrt{\bs{P}_0} \ \text{ (Cholesky factor)}, \quad
                        \bs{L}_X^{0} = \bs{0} \in \mathbb{R}^{r \times
                        p}, \quad \bs{U}^{0} =\bs{P}_\alpha\equiv
                    \alpha\bs{I}_{(*)}(\bs{I}_{(*)})^T}\label{eq:algo-estim-1}
                \end{multline}

                \noindent\hspace{-1.2cm} Then, for $k=1, \cdots, N_T$:
            \item \textbf{Sampling:} generate $2p$ particles from the current state and
                parameter estimates, i.e. for $i=1,\ldots,2p$:
                \begin{multline}
                    \shoveright{\hspace{0.1cm}
                        \begin{cases}
                            \bs{\hat{X}}^{k-1}_{(i)} = \bs{\hat{X}}^{k-1}_+ +
                            \bs{L}_X^{k-1}(\bs{C}^{k-1})^{T}\bs{I}_{(i)}, \\
                            1\leq i \leq p+1 \\
                            \bs{\hat{\theta}}^{k-1}_{(i)} = \bs{\hat{\theta}}^{k-1}_+ +
                            \bs{L}^{k-1}_{\theta}(\bs{C}^{k-1})^{T}\bs{I}_{(i)}
                    \end{cases}}\label{eq:algo-estim-2}
                \end{multline}
                with $\bs{C}^{k-1}$ the lower Cholesky factor of $(\bs{U}^{k-1})^{-1}$.

            \item \textbf{Prediction:} propagate each particle with the forward model
                and compute an a priori state prediction:
                \begin{multline}
                    \shoveright{\hspace{0.1cm}
                        \begin{cases}
                            \bs{\hat{X}}^k_{(i)} =
                            \mathcal{A}^{k}(\bs{\hat{X}}^{k-1}_{(i)},\bs{\hat{\theta}}^{k-1}_{(i)}),\quad
                            \bs{\hat{\theta}}^{k}_{(i)} =\bs{\hat{\theta}}^{k-1}_{(i)},  \quad
                            i=1,\dots,2p                        \\
                            \bs{\hat{X}}^k_- = E_\alpha(\bs{\hat{X}}^{k}_{(*)}) \equiv \alpha
                            \sum_{i=1}^{2p} \bs{\hat{X}}^k_{(i)} \\
                            \bs{\hat{\theta}}_k^- =
                            E_\alpha(\bs{\hat{\theta}}^k_{(*)})
                    \end{cases}}\label{eq:algo-estim-3}
                \end{multline}
            \item \textbf{Innovation computation:} for each particle $j$, compute the innovation
                $\bs{\Gamma}_{(j)}^k$:
                First, compute
                \begin{equation}
                    \bs{\tilde{\Gamma}}_{(j)}^k =
                    \begin{bmatrix}
                        \Re(\bs{Y}^k) -
                        \Re(\mathcal{H}_{\mathcal{F}}(\mathcal{H}(\mathcal{A}^k(\bs{\theta}_{(j)}))))
                        \\
                        \Im(\bs{Y}^k) -
                        \Im(\mathcal{H}_{\mathcal{F}}(\mathcal{H}(\mathcal{A}^k(\bs{\theta}_{(j)}))))
                    \end{bmatrix} \in \mathbb{R}^{2\times N_x \times N_y \times N_z}
                \end{equation}
                and then construct the innovation by excluding all zero values and
                mapping the matrix
                to a vector of equivalent dimension, leading to a vector $\bs{\Gamma}_{(j)}^k \in
                \mathbb{R}^{2N_S}$ where $N_S$ is the number of sampled frequencies.
            \item \textbf{Correction:}  compute a posteriori estimates based on
                measurements for state and parameters, using
                the $i$-th
                particle innovation
                $\bs{\Gamma}^k_{(i)}$:
                \begin{multline}
                    \shoveright{\hspace{0.1cm}
                        \begin{cases}
                            \bs{L}^k_X = \alpha \bs{\hat{X}}^k_{(*)}  (\bs{I}_{(*)})^{T}
                            \\
                            \bs{L}_\theta^k = \alpha \bs{\hat{\theta}}^k_{(*)} (\bs{I}_{(*)})^{T}
                            \\ 
                            \bs{L}_\Gamma^k =  \alpha \bs{\Gamma}^k_{(*)}  (\bs{I}_{(*)})^{T}
                            \\
                            \bs{U}^k =  \bs{P}_\alpha +  (\bs{L}^k_\Gamma)^{T} \bs{W}^{-1}
                            \bs{L}^k_\Gamma    \\
                            \bs{\hat{X}}^k_+ = \bs{\hat{X}}^k_- - \bs{L}^k_X
                            (\bs{U}^k)^{-1}(\bs{L}^k_\Gamma)^{T} \bs{W}^{-1}
                            E_\alpha(\bs{\Gamma}^k_{(*)})    \\
                            \bs{\hat{\theta}}_+^k = \bs{\hat{\theta}}^k_- - \bs{L}^k_\theta
                            (\bs{U}^k)^{-1}(\bs{L}^k_\Gamma)^{T}
                            \bs{W}^{-1}E_\alpha(\bs{\Gamma}^k_{(*)})     \\
                            \bs{P}_{\theta}^{k} = \bs{L}_{\theta}^{k} (\bs{U}^k)^{-1}
                            (\bs{L}_{\theta}^{k})^T \\
                    \end{cases}}\label{eq:algo-estim-4}
                \end{multline}
            \end{subequations}
    \end{itemize}
    with $\bs{W}=\sigma_y^2 \mathbb{I}$.
    \end{document}